\documentclass[letter, traditabstract, longauth]{aa} % for the letters 
\usepackage{graphicx, natbib}
\usepackage{txfonts,latexsym,amssymb}
\begin{document}
   \title{Recent star formation in the Lupus clouds as seen by {\em Herschel}\thanks{{\em Herschel} is an ESA space observatory with science instruments provided by European-led Principal Investigator consortia and with important participation from NASA.}}

   \subtitle{}

   \author{K.~L.~J. Rygl\inst{1}
          \and
          M. Benedettini\inst{1} %\fnmsep\thanks{}
          \and
          E. Schisano\inst{1}
          \and 
          D. Elia\inst{1}
          \and
          S. Molinari\inst{1}
          \and 
          S. Pezzuto\inst{1}
           \and
          Ph. Andr\'e\inst{2}
          \and
          J.~P. Bernard\inst{3}
          \and
          G.~J. White\inst{4,5} 
           \and
          D. Polychroni\inst{6,1}
          \and
          S. Bontemps\inst{7,2} 
          \and
          N.~L.~J. Cox\inst{8}
          \and
          J. Di Francesco\inst{9,10}
          \and
          A. Facchini\inst{1}
          \and
          C. Fallscheer\inst{9,10}
          \and
          A.~M. di Giorgio\inst{1}
          \and
          M. Hennemann\inst{2}
          \and
          T. Hill\inst{2}
          \and
          V. K\"onyves\inst{2}
          \and
          V. Minier\inst{2}
          \and
          F. Motte\inst{2}
          \and
          Q. Nguyen-Luong\inst{11}
          \and
          N. Peretto\inst{2}
          \and
          M. Pestalozzi\inst{1}
          \and
          S. Sadavoy\inst{9,10}
          \and
          N. Schneider\inst{7,2}
          \and
          L. Spinoglio\inst{1} 
          \and 
          L. Testi\inst{12,13} 
          \and
          D. Ward-Thompson\inst{14}        
         }
   \institute{Istituto di Astrofisica e Planetologia Spaziali (INAF-IAPS), 
   via del Fosso del Cavaliere 100, 00133 Roma, Italy\\ 
         \email {kazi.rygl@inaf.it}    
         \and
         Laboratoire AIM Paris-Saclay, CEA/IRFU CNRS/INSU Universit\'e Paris Diderot, 91191 Gif-sur-Yvette, France
         \and
         CESR, Observatoire Midi-PyrŽnŽes (CNRS-UPS), Universit\'e de Toulouse, BP 44346, 31028 Toulouse, France
         \and
         Rutherford Appleton LIbrary, Chilton, Didcot, OX11 0NL, UK
         \and
         Department of Physics and Astronomy, Open University, Milton Keynes, UK
         \and
         University of Athens, Department of Astrophysics, Astronomy and Mechanics, Faculty of Physics, Panepistimiopolis, 15784 Zografos, Athens, Greece
         \and
         CNRS/INSU, Laboratoire d'Astrophysique de Bordeaux UMR 5904, BP 89, 33271 Floirac, France
          \and
         Instituut voor Sterrenkunde, KU Leuven, Celestijnenlaan 200D, 3001 Leuven, Belgium
         \and
         Department of Physics and Astronomy, University of Victoria, PO Box 355, STN CSC, Victoria BC Canada, V8W 3P6
         \and
         National Research Council Canada, Herzberg Institute of Astrophysics, 5071 West Saanich Road, Victoria BC Canada, V9E 2E7
         \and
         Canadian Institute for Theoretical Astrophysics (CITA), University of Toronto, 60 St. George Street, Toronto ON Canada, M5S 3H8
         \and
         European Southern Observatory, Karl-Schwarzschild-Strasse 2, 87548 Garching bei M\"unchen, Germany
         \and
         INAF-Osservatorio Astrofisico di Arcetri, Large E. Fermi 5, 50125 Firenze, Italy
         \and
         Jeremiah Horrocks Institute, University of Central Lancashire, PR1 2HE, UK}             

   \date{Received ; Accepted}

% \abstract{}{}{}{}{} 
% 5 {} token are mandatory
 
  \abstract
{We present a study of the star formation histories of the Lupus\,I, III, and IV clouds using the {\em Herschel} 70--500$\,\mu$m maps obtained by the {\em Herschel} Gould Belt Survey Key Project. By combining the new {\em Herschel}  data with the existing {\em Spitzer} catalog we obtained an unprecedented census of prestellar sources and young stellar objects in the Lupus clouds, which allowed us to study the overall star formation rate (SFR) and efficiency (SFE).
The high SFE of Lupus\,III, its decreasing SFR, and its large number of pre-main sequence stars with respect to proto- and prestellar sources, suggest that Lupus\,III is the most evolved cloud, and after having experienced a major star formation event in the past, is now approaching the end of its current star-forming cycle.
Lupus\,I is currently undergoing a large star formation event, apparent by the increasing SFR, the large number of prestellar objects with respect to more evolved objects, and the high percentage of material at high extinction (e.g., above $A_V\approx8\,$mag). Also Lupus\,IV has an increasing SFR; however, the relative number of prestellar sources is much lower, suggesting that its star formation has not yet reached its peak. 
}


 \maketitle
%
%________________________________________________________________

\section{Introduction}
In the current paradigm of low-mass star formation (SF), a gravitationally bound prestellar core will evolve into a  young stellar object (YSO), passing through several phases, usually defined by Classes representing increasing stages of evolution: 0, I, II, and III (see \citealt{andre:2000,lada:1984b} for the definitions), before becoming a main-sequence star. While the later stages of low-mass SF are largely understood, less is known about the earlier stages (including the prestellar cores and the Class 0 objects) due to a lack of sensitivity and resolution at far-infrared to submm wavelengths. The {\em Herschel} Gould Belt Survey (HGBS, \citealt{andre:2010}), carried out with the {\em Herschel} Space Observatory (\citealt{pilbratt:2010}), aims at studying these early stages of SF in nearby molecular clouds forming the so-called Gould Belt (\citealt{comeron:1992}).

Located at a distance between 150\,pc (Lup\,I and IV) and 200\,pc (Lup\,III; \citealt{comeron:2008}),
the Lupus clouds (I, III, IV) are among the nearest star-forming regions in the Gould Belt. 
The large angular extent of the Lupus clouds across the sky (334$^\circ$$<$$l$$<$352$^\circ$, 5$^\circ$$<$$b$$<$25$^\circ$) corresponds to a physical extent of 50\,pc$\times$55\,pc at a distance of 150\,pc, similar to the distance range among the Lupus \,I, III, and IV, clouds (50\,pc). 
Previous  {\em Spitzer} (\citealt{merin:2008}, hereafter M08) and molecular-line (\citealt{benedettini:2011}) studies of Lupus\,I, III, and IV have found that the three clouds seem to be at different stages of evolution: Lupus I is thought to be the youngest cloud, Lupus IV is a little more evolved, and Lupus III is the most evolved cloud. 

The Lupus\,I, III, and IV clouds were mapped, as a part of the HGBS, at five wavelengths from 70\,$\mu$m to 500\,$\mu$m, covering the range where the spectral energy distribution (SED) of cold dust emission from prestellar sources and envelopes of Class 0/I objects (protostars), is likely to peak. Therefore, the {\em Herschel} data are crucial for detecting these objects and determining their physical parameters. The combination of the M08 catalog, containing mostly Class II/III pre-main sequence (PMS) stars, with the prestellar cores and Class 0/I sources detected by {\em Herschel} allows for a much more complete view of the ongoing SF in the Lupus clouds than previously possible. Here, we present the first-look letter of the SF history of the Lupus clouds. A detailed analysis of the {\em Herschel} data on Lupus and the resulting catalog of identified objects will be presented in a forthcoming first-generation paper (Benedettini et al.~in prep.) and will be publicly available at the HBGS website\footnote{http://gouldbelt-herschel.cea.fr/archives}.

\section{{\em Herschel} observations and data reduction}
The Lupus maps were obtained between January 2010 and January 2011 by photometric observations with the Photodetector Array Camera and Spectrometer (PACS; \citealt{poglitsch:2010}) and  Spectral and Photometric Imaging Receiver (SPIRE; \citealt{griffin:2010}) in parallel mode using a scanning speed of 60\arcsec s$^{-1}$. Map sizes are $2^\circ$$\times$$2\rlap{.}^\circ$3 for Lupus\,I and $1\rlap{.}^\circ 5$$\times$$1\rlap{.}^\circ 1$ for Lupus\,III, covering a similar region as the {\em Spitzer} observations for both clouds. Lupus\,IV was imaged in two maps, $2^\circ $$\times$$ 1\rlap{.}^\circ3$ and $1\rlap{.}^\circ3$$\times$$ 1\rlap{.}^\circ3$, covering the {\em Spitzer} map, as well as a new region, never mapped before at wavelengths of 160--500\,$\mu$m. 
The data were exported from HIPE v\,7.0 (\citealt{ott:2010}) at level 0.5, and processed with the ROMAGAL data reduction pipeline (\citealt{traficante:2011,piazzo:2012}). 
The maps were astrometrically aligned with the 70\,$\mu$m MIPS images from the {\em Spitzer} `cores2disks' (c2d) survey (\citealt{evans:2003b}),  which in turn have been aligned with 2MASS data, based on a number of point sources observed in both 70\,$\mu$m maps and yielding an astrometric precision of $\sim$$2\rlap{.}\arcsec5$. Absolute flux calibration (see \citealt{bernard:2010}) was found to be better than 20\% by comparing the {\em Herschel} data with the {\em Planck} and {\em IRAS} data in the same area. 
The resulting maps have beam sizes of 9\arcsec, 12\arcsec, 18\arcsec, 25\arcsec, 36\arcsec, for 70\,$\mu$m, 160\,$\mu$m, 250\,$\mu$m, 350\,$\mu$m, and 500\,$\mu$m, respectively. 
The Lupus\,I observations were affected by stray Moonlight, visible as a bright band in the declination direction of the map. Fortunately, this did not affect our compact source fluxes, since the background is removed in the procedure.

For computing column density and temperature maps, the 70--350\,$\mu$m maps were convolved to the 500\,$\mu$m resolution and rebinned to the same pixel size (11$\rlap{.}$\arcsec 5). The pixel-by-pixel modified black-body fits were then performed on the regridded maps, excluding the 70\,$\mu$m map since this emission might not be tracing the cold dust exactly. For the modified black-body fitting we assumed a dust opacity of $\kappa_{1.3\,\mathrm{mm}}$=$0.004\,\mathrm{cm^2 g^{-1}}$ (\citealt{hildebrand:1983}), a grain emissivity parameter $\beta$=2, and the mean molecular weight $\mu$=2.8 (\citealt{kauffmann:2008}).
The resulting column density map (Fig.\,\ref{fig:maps}) was used to define the cloud area: the emission within the $A_V$$=$2\,mag (assuming the column density to $A_V$ relation of $N_{\mathrm{H_2}}$=9.4$\times10^{20}\,A_V\,\mathrm{cm^{-2}}$, \citealt{bohlin:1978}) contour was considered cloud emission and the integrated column densities in this contour (Table \ref{ta:one})
agree within 20\% with the masses from the c2d extinction maps\footnote{http://data.spitzer.caltech.edu/popular/c2d/20071101\_enhanced\_v1/} (\citealt{chapman:2009}) within the same area.

Compact source detection and extraction were performed with CuTEx (\citealt{molinari:2011}) in each of the five {\em Herschel} maps separately, using a 3$\sigma$ SNR detection limit. Following \citet{elia:2010}, sources across the five bands were associated according to their positions. We adopted a conservative approach and removed sources that had a displaced counterpart in the 350\,$\mu$m or 500\,$\mu$m bands by more than half of the FWHP from the source center common to the other bands, which can introduce a large uncertainty in  the measured fluxes at the longer wavelengths.  
In particular, for the prestellar cores this could result in finding false objects by an overestimation of the mass (see next section).
Sources with detection in fewer than three bands longward of $\lambda$=70$\,\mu$m were discarded. 
We then fit a single temperature modified black-body function to the SEDs of the extracted sources, again excluding the $70\,\mu$m flux, to derive dust temperatures. Source masses were then derived from the optically thin part of the modified black-body spectrum, using the same dust properties as above. The reduced chi-squared was used to remove sources with badly fitted SEDs, which would yield uncertain masses and temperatures otherwise, and hence could have a faulty prestellar core classification. 
Objects identified as non-Lupus members based on their proper motions (\citealt{lopez:2011}), in total 16 objects, were removed. 
%Galaxies can contaminate the prestellar and YSO sample. 
We cross-checked our sample with the SIMBAD and the c2d database to remove known extragalactic sources (in total 13 galaxies were found). Seven off-cloud unresolved objects without a 70$\,\mu$m counterpart were classified as candidate galaxies and removed. %(not shown in Fig.~\ref{fig:maps}). 
Finally, to understand whether some of the off-cloud candidate protostars might have been misclassified (see Fig.\,\ref{fig:maps} and Sect. \ref{sec:disc}),
we estimated the galaxy count completeness. Using the number counts from \citet{clements:2010}, we find that for Lupus\,I and III we start lacking galaxies with flux levels of $\sim$350\,mJy at 250\,$\mu$m, while for Lupus\,IV the limit is 700\,mJy.

\section{Results}

\begin{table} [!ht]
\caption{Properties of the Lupus clouds\label{ta:one}}
\begin{tabular}{l c c c}
\hline\hline
Cloud & Lupus\,I & Lupus\,III & Lupus\,IV\\
\hline
distance\tablefootmark{a}  (pc) & 150 & 200 & 150\\
coverage (degree$^2$)\tablefootmark{b}  &4.6 & 1.6&4.3\\
cloud area$> $$A_V$=2\,mag (pc$^2$) & 16.4 & 7.7 & 7.1 \\
cloud mass$>$$A_V$=2\,mag ($M_\odot$) & 830 & 570 & 500\\ 
cloud mass$>$$A_V$=8\,mag ($M_\odot$) & 145 & 65 & 60\\
$N_\mathrm{tot}=N_\mathrm{YSO}+N_\mathrm{prestellar}$  & 52 & 113 & 37 \\
$N_\mathrm{YSO}$ 0/I/II/III\tablefootmark{c} &1/10/10/4 & 1/10/50/42 & 3/9/11/4\\ 
$N_\mathrm{prestellar}$ &27 & 10 & 10\\ 
$N_\mathrm{unbound}$ & 68 &8 & 63 \\
\hline
\end{tabular}
\tablefoot{
\tablefoottext{a}{See \citet{comeron:2008}}
\tablefoottext{b}{SPIRE and PACS overlapping area.}
\tablefoottext{c}{Number of objects per Class 0, I, II, and III are based on the {\em Herschel} data and on the M08 catalog.}}
\end{table}

\begin{figure}
\includegraphics[width=11.0cm,angle=-90]{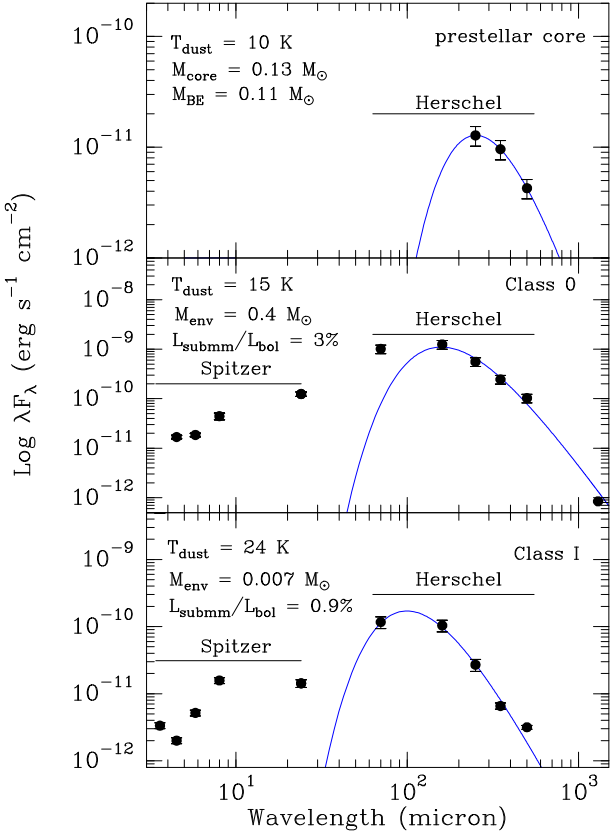}
\caption{\label{fig:sed} Spectral energy distributions of a prestellar core, Class 0, and a Class I source in Lupus I and the modified black-body fit to the cold dust emission (blue line). The 3.4--24\,$\mu$m  and 1300\,$\mu$m data points were taken from M08.}
\end{figure}

Figure \ref{fig:maps} shows that the Lupus clouds are overall elongated in shape and fragmented into smaller clumps, in which most of the pre- and protostellar objects are located. A prestellar core is defined as a gravitationally bound dense core (size $<0.05$\,pc) without an internal luminosity source that will supposedly form stars in the future (see e.g., \citealt{ward:1994,difrancesco:2007}). We used the ratio of the core mass to its critical Bonnor-Ebert mass $\frac{M}{M_\mathrm{BE}}$, to distinguish between prestellar cores and unbound objects that may or may not form stars (cf. \citealt{konyves:2010}), where $M$ is the core mass, $M_\mathrm{BE}$=$2.4\,R\,c_s^2\,/G$ the Bonnor-Ebert mass (\citealt{bonnor:1956}), $c_s$ the sound speed ($c_s$=$\sqrt{k_B\,T/\mu m_{H}}$), $R$ the radius, and $T$ the core temperature. For the core mass, we used the mass obtained from the modified black-body fitting and computed the $M_{BE}$ using the fitted modified black-body temperature and the deconvolved radius at 250\,$\mu$m.
Cores with  $\frac{M}{M_\mathrm{BE}}>1.0$ are gravitationally bound (\citealt{bonnor:1956}) and were classified as prestellar, while cores with a lower ratio were considered unbound objects. An example of prestellar core SED is given in Fig.~\ref{fig:sed}.
 
A protostar is defined as a core that is internally heated by an embedded YSO and is still accreting material from its envelope. The presence of a YSO is ascertained by a 70\,$\mu$m and/or 24\,$\mu$m source (\citealt{konyves:2010}). Among the protostellar objects, the submm to bolometric luminosity ratio (\citealt{andre:2000}), $L_{\mathrm{s2b}}=L_\mathrm{submm}^{\lambda\geq350\mu\mathrm{m}}/L_\mathrm{bol} ^{\lambda\geq3.4\mu\mathrm{m}}$, derived from the {\em Herschel}, M08, and WISE data (\citealt{wrightwise:2010}), was used to identify the protostars as Class 0 or Class I objects. Protostars with $L_{\mathrm{s2b}}$$\geq$3\% 
were classified as candidate Class 0, while protostars with a lower ratio were identified as candidate Class I (a more conservative Class 0 definition than used by \citealt{bontemps:2010}). Examples of a Class 0 and a Class I SED are shown in Fig.~\ref{fig:sed}. The classification of more evolved objects, i.e., the PMS stars (or Class II/III objects), was taken from M08 and SIMBAD. M08 also identified flat SED objects that trace the transition between Class I and II (\citealt{greene:1994}). Since we cannot distinguish between Class I and F sources with the {\em Herschel} data, we considered them as Class I objects. 

\begin{figure*}
\centering
\includegraphics[angle=-90,width=9cm]{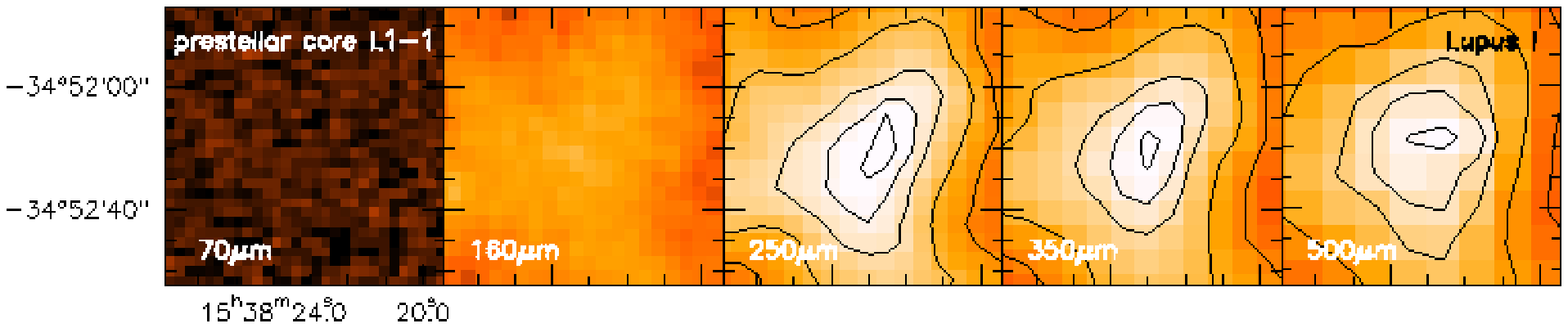}
\includegraphics[angle=-90,width=9cm]{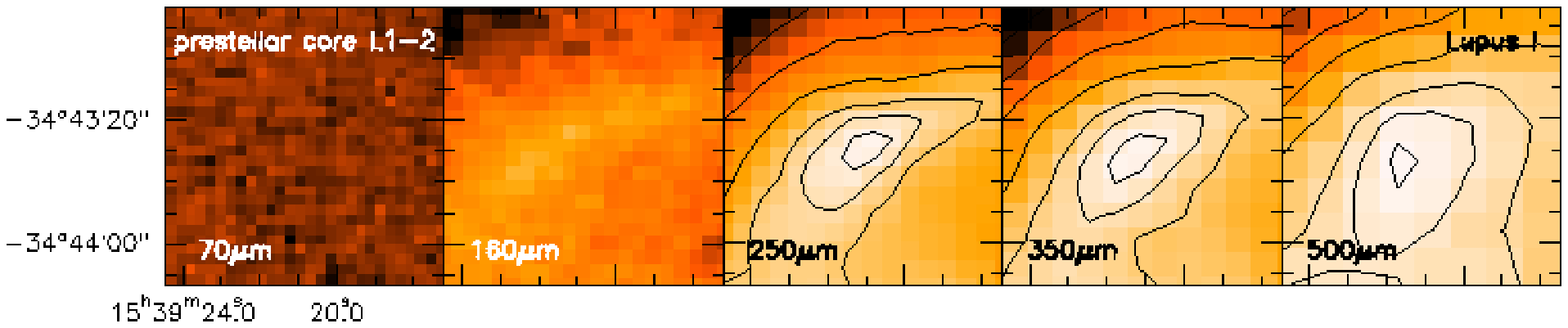}
\includegraphics[angle=-90,width=9cm]{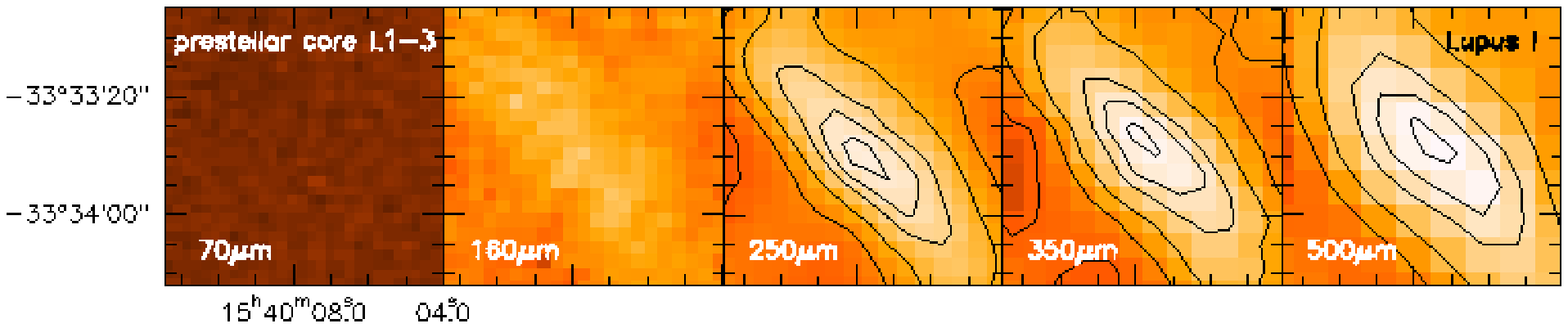}
\includegraphics[angle=-90,width=9cm]{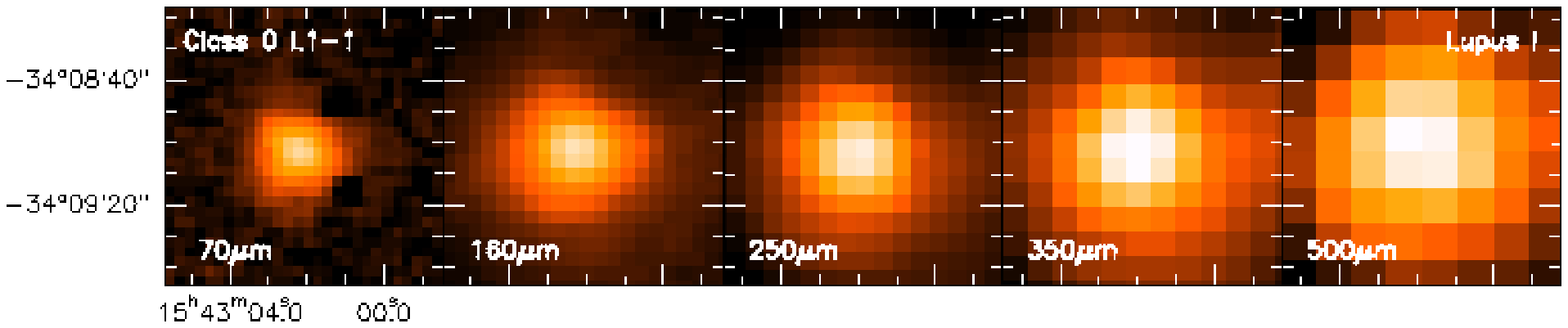}
\caption{\label{fig:map5} Five-band images of three prestellar cores (contour levels are peakflux$\ast$0.99, 0.95, 0.90, 0.80, 0.70, 0.60, and 0.50) and a Class 0 object in Lupus I. The maps are centered on the prestellar core/Class 0 object and are ordered by right ascension. The remaining prestellar and Class 0 sources of all three Lupus clouds are given in Fig.\,\ref{fig:youngsources}.}
\end{figure*}

With the conservative source selection (Sect. 2) and the classification described above, we found 47 candidate prestellar cores (see Fig.~\ref{fig:map5}, continued in Fig.\,\ref{fig:youngsources}, for all the five-band images). For the entire Lupus complex, the prestellar cores have median masses of 0.25$\,M_\odot$ within a range of 0.1--3.0\,$M_\odot$, and median temperatures of 9\,K within a range of 7--14\,K. The mean prestellar source size is $0.02\pm0.01$\,pc. 
Five Class 0 sources were found (Figs.~\ref{fig:map5} and \ref{fig:youngsources}), of which most were previously classified as Class I objects (M08). The only previously known Class 0 is Lup\,MM3 (\citealt{tachihara:2007}). The Class 0 temperatures are between 9.5\,K and 15\,K. Their envelope masses ranged from 0.08 to 0.6\,$M_\odot$. For two Class 0 sources, we compared the dust temperatures obtained from the {\em Herschel} data to the kinetic temperatures obtained from ammonia (\citealt{benedettini:2011}), and found that they agree within 20\%.  
We found 14 new Class I candidates in the Lupus clouds, 11 of which were outside the field covered by {\em Spitzer} c2d. 

Table \ref{ta:one} summarizes the properties of the Lupus clouds and the objects found therein based on our results and the M08 catalog. 
The overlap between the {\em Herschel} data, including the YSOs  that were only detected at $\lambda$=70\,$\mu$m and 160$\,\mu$m, with the M08 catalog was quite large, especially for the younger Class I objects. We detected 
61\% of their Class I objects, 46\% of their Class II objects, and 10\% of their Class III objects in the area covered by both surveys. For the few undetected Class I objects, we place an upper limit of 66 mJy at 250\,$\mu$m, and assuming a point source morphology, this implies a mass limit of $0.002\,M_\odot$ at a dust temperature of 15\,K.

 \section{Star formation history}
\label{sec:disc}
\citet{palla:2000} posit that the star formation rate (SFR) has been increasing in the Lupus clouds over the past 4\,Myr using number counts of PMS stars, whose ages they estimated through comparison with theoretical evolutionary PMS tracks. Our sample contains both prestellar cores and Class 0/I objects, and by merging with the M08 catalog we can, therefore, estimate the recent SFR behavior better than in previous studies. 

\begin{figure}
\centering
\includegraphics[angle=-90,width=8.8cm]{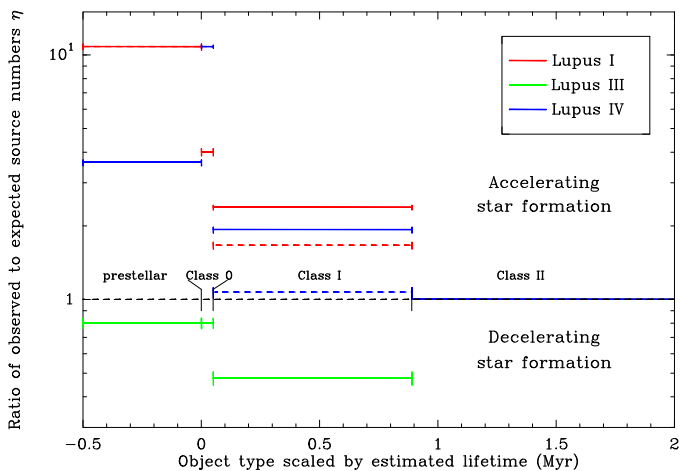} 
\caption{\label{fig:hist} Ratio of observed-to-expected (for a constant SFR) source numbers, $\eta$, per Class. Solid lines represent the numbers from Table \ref{ta:one}, while dashed lines show Class I without the possible galaxy contamination. }
\end{figure}

One can deduce the relative number of objects $N$ of a certain class with respect to a reference class expected for a constant SFR as $N$=$N_\mathrm{ref}$$\times$$\tau/\tau_\mathrm{ref}$ by assuming the classes' lifetimes $\tau$ and $\tau_\mathrm{ref}$. Since the Class II  objects are expected to be complete (M08 and SIMBAD), we chose this as the reference class with the lifetime estimate of 2$\pm$1\,Myr (see \citealt{evans:2009}).
For the other classes, we assume 0.5\,Myr for the prestellar lifetime (\citealt{enoch:2008}), 0.05\,Myr for the Class 0 lifetime (\citealt{froebrich:2006b}), and 0.84\,Myr for the Class I lifetime (\citealt{evans:2009}). 
Using these lifetimes, we derived the expected numbers of objects in each Class (relative to the number of Class II objects). 
Figure~\ref{fig:hist} shows the ratio $\eta$ of the observed-to-expected source numbers for the prestellar ($\eta_\mathrm{pres}$), Class 0 ($\eta_\mathrm{Cl\,0}$), and I ($\eta_\mathrm{Cl\,I}$) objects in each cloud. In this figure, we also plot the $\eta_\mathrm{Cl\,I}$ without the possible galaxy contaminated sources (see Fig.\,\ref{fig:maps}) based on the estimated completeness limit. Clearly, this possible contamination does not influence our SFR conclusions. 

For both Lupus\,I  and IV we find more prestellar, Class 0, and Class I objects than predicted for a constant SFR. We are therefore witnessing an increasing SFR over the past 0.5--1.5 Myr. The $\eta$'s,  particularly the $\eta_\mathrm{pres}$, is much larger in Lupus\,I than in Lupus\,IV, suggesting that Lupus\,I is undergoing a star formation event, while the SFR of Lupus IV might increase even more in the future.
In Lupus\,III there are fewer prestellar, Class 0, and Class I objects than expected for constant SFR. The star formation in Lupus\,III has decelerated over the past 2\,Myr. 

The SFR analysis depends heavily on the lifetimes used in the estimation. Performing the inverse analysis, assuming a constant SFR, and deriving the lifetimes (with respect to the number of Class II objects) would lead to lifetimes longer by a factor $\sim$10 for the prestellar sources, $\sim$4 for the Class 0 objects, and $\sim$2 for the Class I objects in Lupus I.  Finally, the results of this analysis do not depend on the source-finding algorithm, CuTEx, since analysis of the sources found by {\em getsources} (\citealt{mensh:2012}) yielded similar SFR trends.   

We calculated the star formation efficiency (SFE), which is the ratio of the total mass in YSOs (Class 0 -- III) $M_s$, assuming 0.2\,$M_\odot$ as the average Lupus PMS star mass (M08), to the total mass of the cloud plus YSOs: $\frac{M_s}{M_{cloud}+M_s}$. 
With this formulation, we find a similar SFE of $\sim$1\% in Lupus\,I and IV, but a much higher SFE of 3.5\%in Lupus\,III. The latter may be expected from the high YSO-to-prestellar objects ratio (Table~\ref{ta:one}). 
The different distance of Lupus\,III, 200\,pc, does not influence these results, because putting Lupus\,III at 150\,pc magnifies its SFE to 6.1\%.

One can study the star formation histories by interpreting the SFR behavior and SFE of the clouds as different evolutionary states within a star formation cycle. 
We propose that the high SFE of Lupus\,III, its decreasing SFR, and large number of PMS stars with respect to proto- and prestellar sources suggest that Lupus\,III is the most evolved cloud, which after having experienced a major star formation event, is now approaching the end of its current star-forming cycle. The properties of Lupus\,III seem to be similar to those of Chameleon I, where \citet{belloche:2011a} claim to see the end of star formation based on prestellar cores found in 870\,$\mu$m LABOCA data.
On the other hand, Lupus\,I is currently undergoing a large star formation event, apparent by the increasing SFR and the large number of prestellar objects with respect to more evolved sources. Also Lupus\,IV has an increasing SFR; however, the relative number of prestellar sources is much lower, suggesting that its star formation has not yet reached its peak and that Lupus\,IV is at an earlier stage of evolution than Lupus I. The contrast between the increasing SFR in Lupus I and the decreasing SFR
in Lupus III is possibly reminiscent of the contrast between L1688 and L1689
in Ophiuchus. There, L\,1688 is an active star-forming cloud with many prestellar cores, while L\,1689 contains only a few prestellar cores, even though these clouds have similar CO properties. However, while \citet{nutter:2006} explain this difference by an external trigger from a nearby OB association, we claim that the diverse SFRs and SFEs in Lupus result from the clouds being in different states of their star formation cycle.

In Table \ref{ta:one} we also list the fraction of cloud mass above the SF threshold of $A_V$$\approx$8\,mag found in recent studies (e.g., \citealt{heiderman:2010,andre:2010}). This number, which is independent of source selection
and identification, provides an estimate of the percentage of cloud mass
directly available for SF, and it is expected to scale as the probability
of finding prestellar cores. \citet{spezzi:2011} were the first to use this method on the extinction maps of the Lupus clouds.  We applied this method to the {\em Herschel} data and found that 
Lupus I has a much higher percentage of mass above the $A_V$ threshold than the Lupus III and IV clouds, supporting the idea of a star formation event in this cloud. Lupus IV has a only slightly higher percentage of dense material than Lupus III, as was also noted by \citet{spezzi:2011}, supporting the similar number of prestellar cores found in these two clouds. The fraction of cloud mass above this $A_V$ threshold correlates well with the number of prestellar cores found with {\em Herschel}, thereby strengthening our conclusions.

\begin{acknowledgements}
The authors thank the anonymous referee for her/his comments. 
SPIRE has been developed by a consortium of institutes led
by Cardiff Univ. (UK) and including: Univ. Lethbridge (Canada);
NAOC (China); CEA, LAM (France); IFSI, Univ. Padua (Italy);
IAC (Spain); Stockholm Observatory (Sweden); Imperial College
London, RAL, UCL-MSSL, UKATC, Univ. Sussex (UK); and Caltech,
JPL, NHSC, Univ. Colorado (USA). This development has been
supported by national funding agencies: CSA (Canada); NAOC
(China); CEA, CNES, CNRS (France); ASI (Italy); MCINN (Spain);
SNSB (Sweden); STFC, UKSA (UK); and NASA (USA).
PACS has been developed by a consortium of institutes led by MPE (Germany) and including UVIE (Austria); KU Leuven, CSL, IMEC (Belgium); CEA, LAM (France); MPIA (Germany); INAF-IFSI/OAA/OAP/OAT, LENS, SISSA (Italy); IAC (Spain). This development has been supported by the funding agencies BMVIT (Austria), ESA-PRODEX (Belgium), CEA/CNES (France), DLR (Germany), ASI/INAF (Italy), and CICYT/MCYT (Spain). KLJR, ES, DE, MP, and DP are funded by an ASI fellowship under contract numbers I/005/11/0 and I/038/08/0. NLJC is supported by the Belgian Federal Science Policy Office via ESA's PRODEX Program.
\end{acknowledgements}

\bibliographystyle{aa} 
\bibliography{/Users/kazi/tot}

\appendix
\section{Online Material}

\onlfig{1}{
\begin{figure*}\centering
\includegraphics[angle=-90,width=14cm]{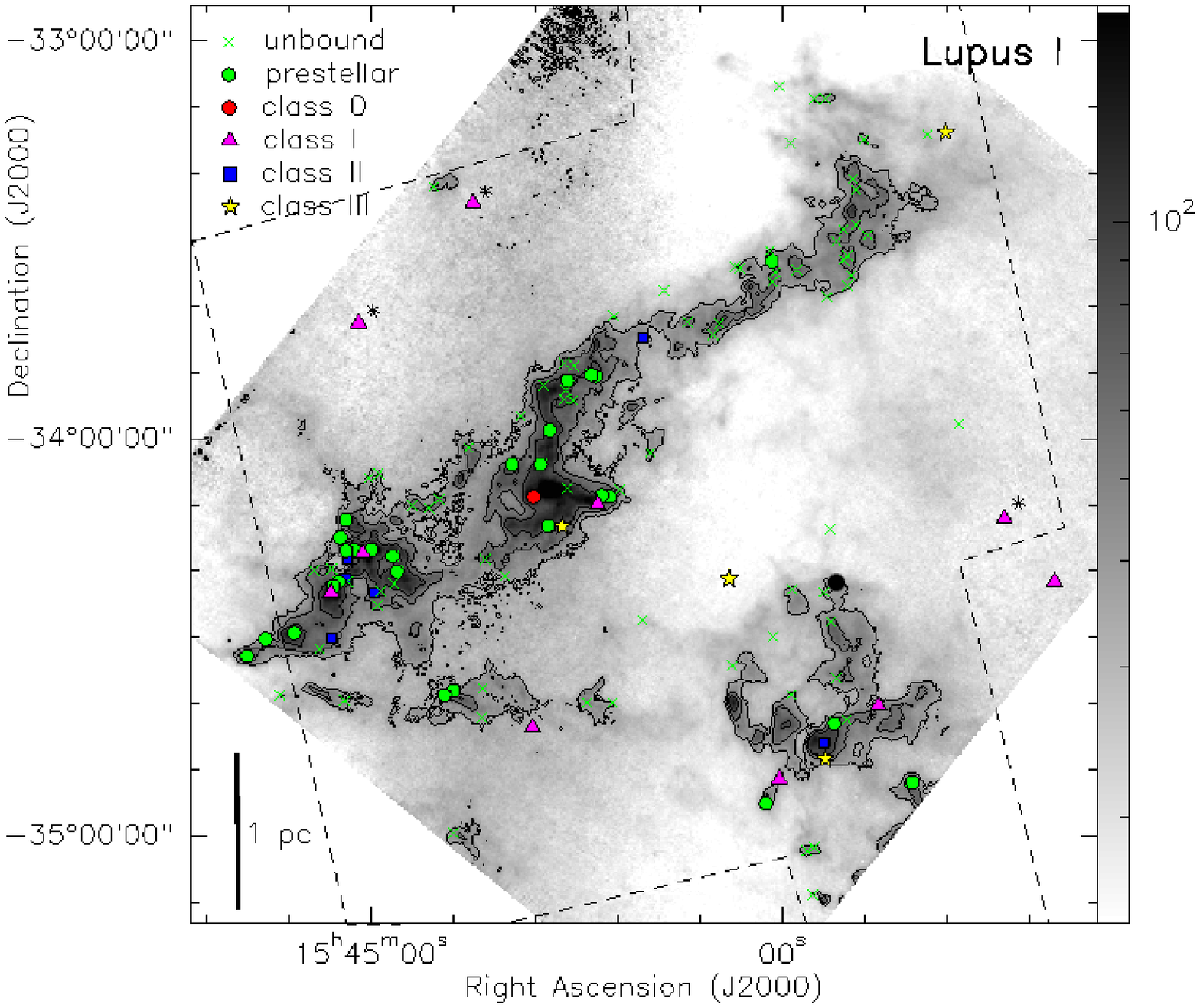}
\includegraphics[angle=-90,width=14cm]{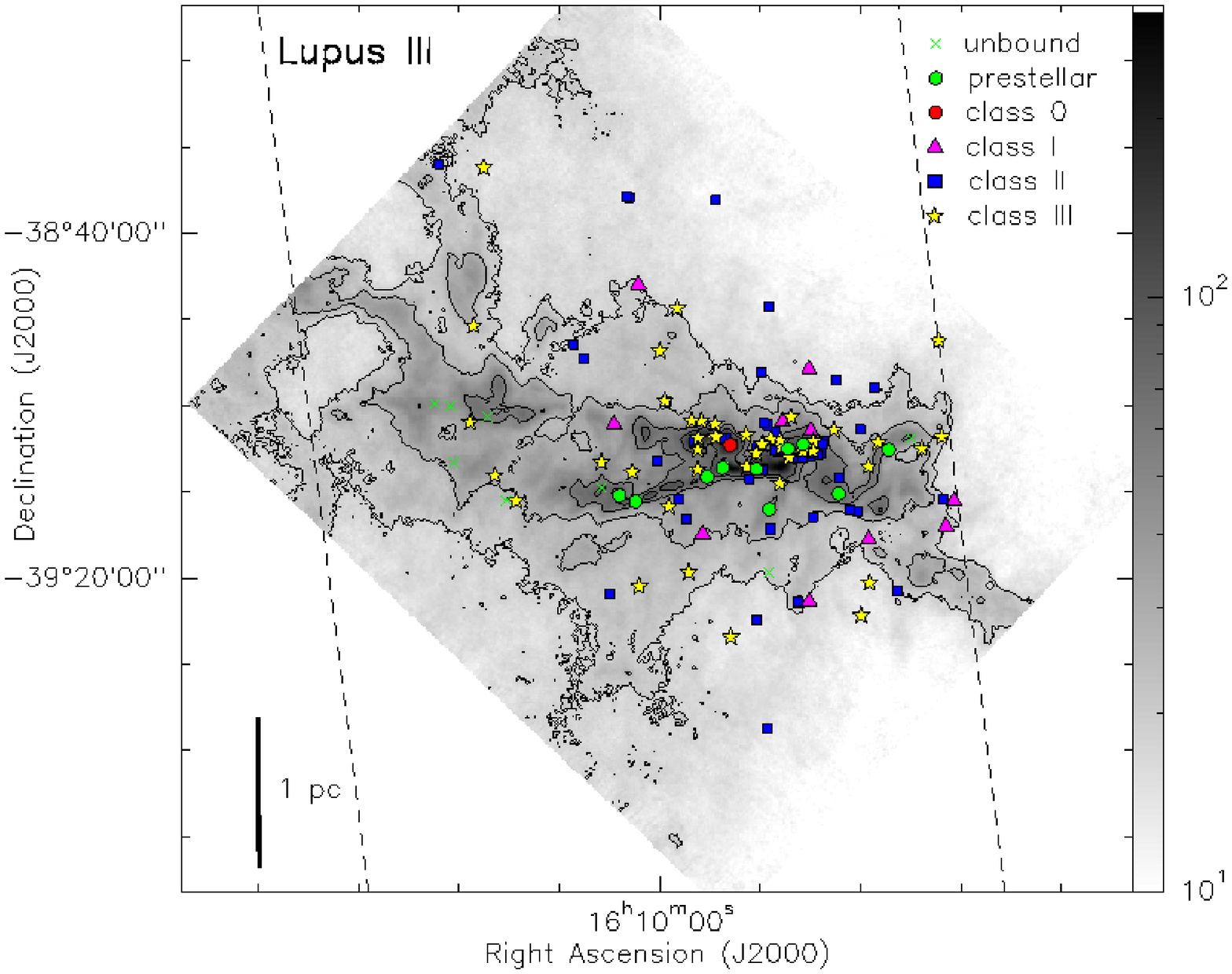}
\caption{ \label{fig:maps} $\mathrm{H_2}$ column density maps of Lupus\,I, III, and IV in units of $10^{20}\,\mathrm{cm^{-2}}$ with $A_V$ contours overplotted. For Lupus\,III and IV the contours are $A_V=$ 2, 3, 6, and 9\,mag, while for Lupus\,I the contours are 4, 6, and 9\,mag (to avoid the stray Moonlight). The different classes of objects, from the {\em Herschel} data and from the M08 catalog, are indicated. The off-cloud candidate Class\,I sources, marked by an asterisk, are those we considered as possible galaxy contaminations (shown by a dashed line in Fig.~\ref{fig:hist}). The one off-cloud Class I in Lupus I without an asterisk has strong 24\,$\mu$m emission, so is less likely to be a galaxy. Dashed contours mark the {\em Spitzer} MIPS coverage. }
\end{figure*}
\addtocounter{figure}{-1}
\begin{figure*}
\centering
%\begin{center}
\includegraphics[angle=-90,width=19cm]{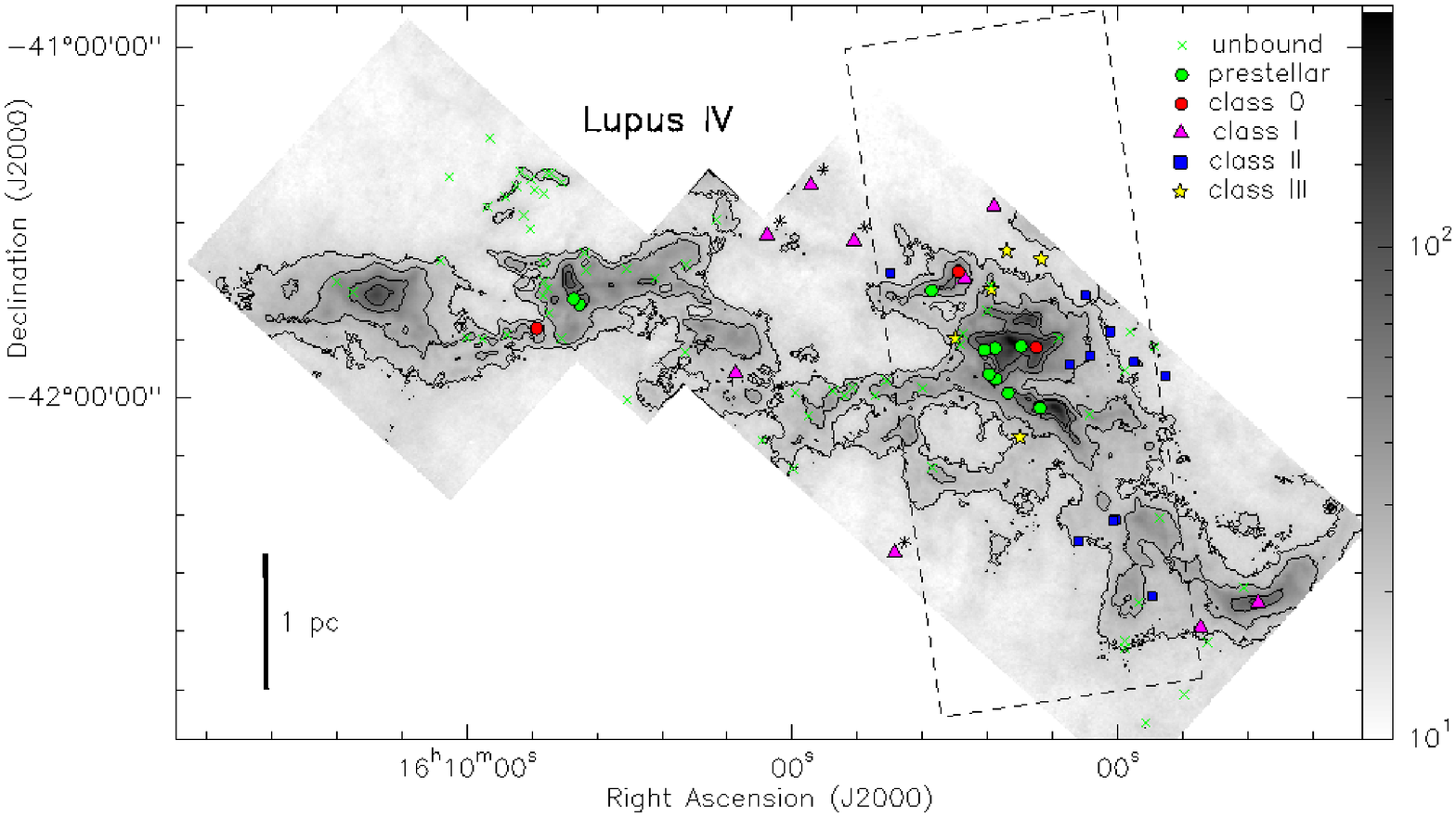}
%\end{center}
\caption{-- {\em Continued}}
\end{figure*}
}

\onlfig{2}{

\begin{figure*}\centering
\includegraphics[angle=-90,width=9cm]{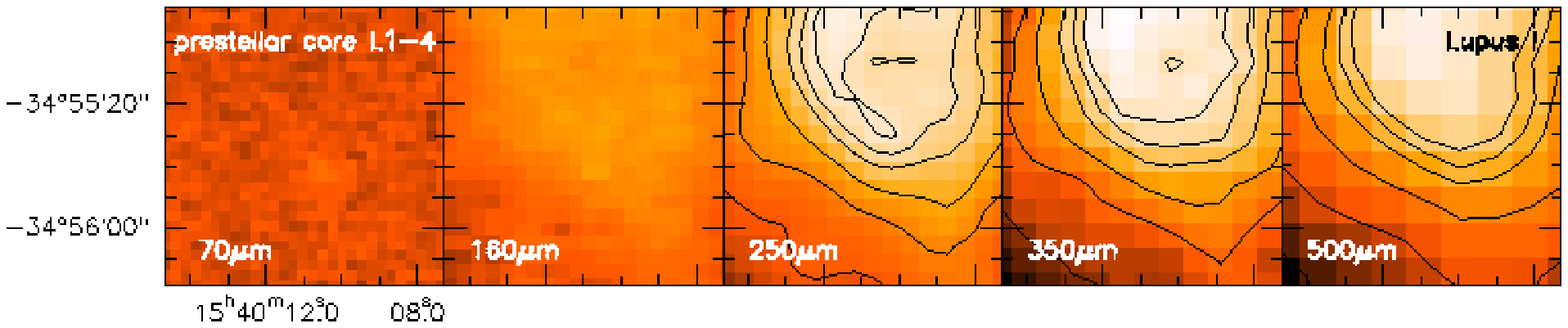}
\includegraphics[angle=-90,width=9cm]{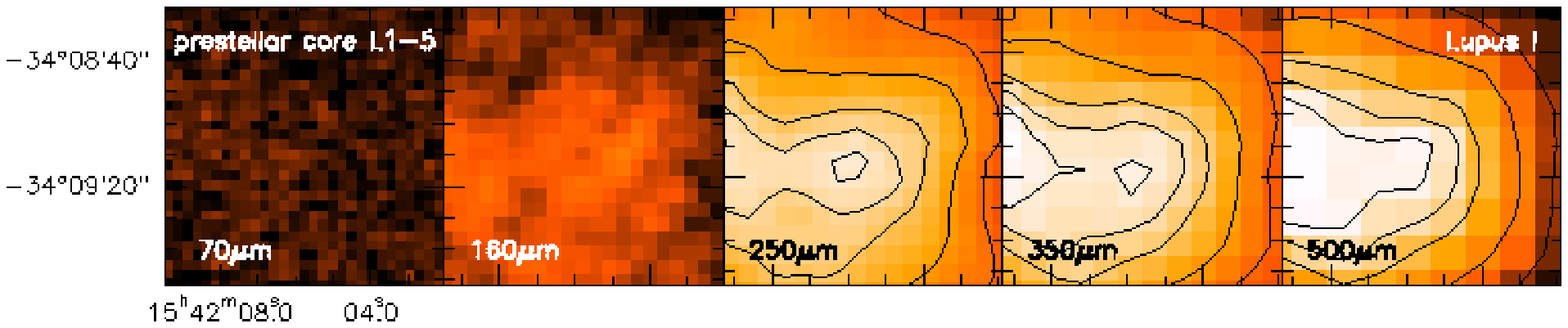}
\includegraphics[angle=-90,width=9cm]{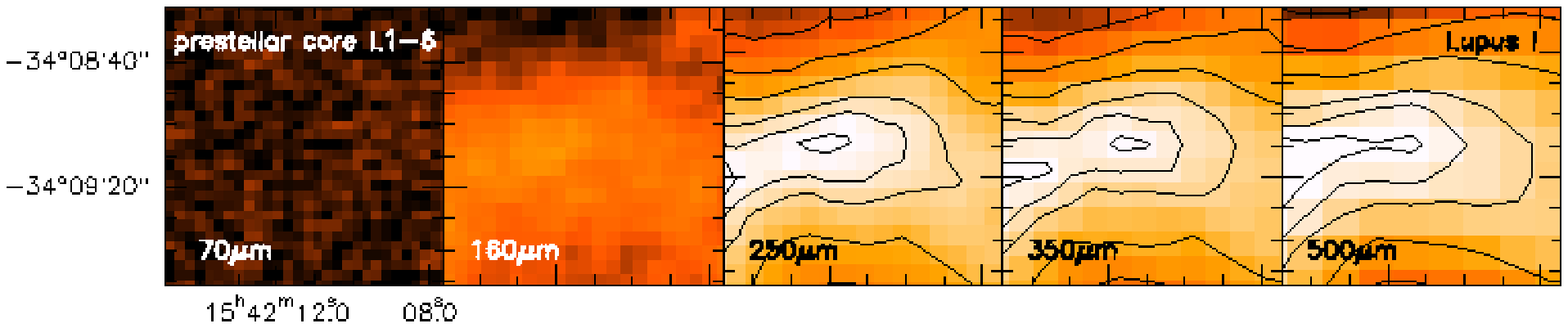}
\includegraphics[angle=-90,width=9cm]{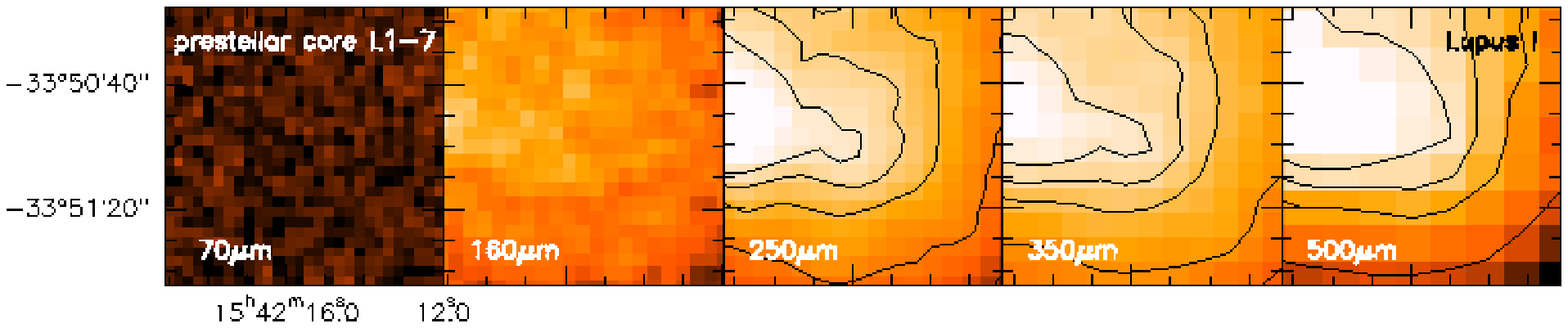}
\includegraphics[angle=-90,width=9cm]{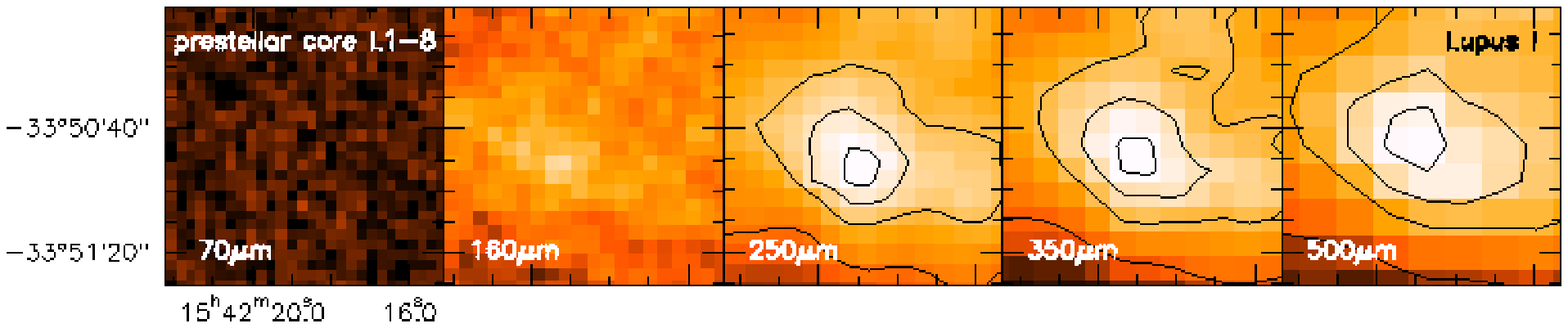}
\includegraphics[angle=-90,width=9cm]{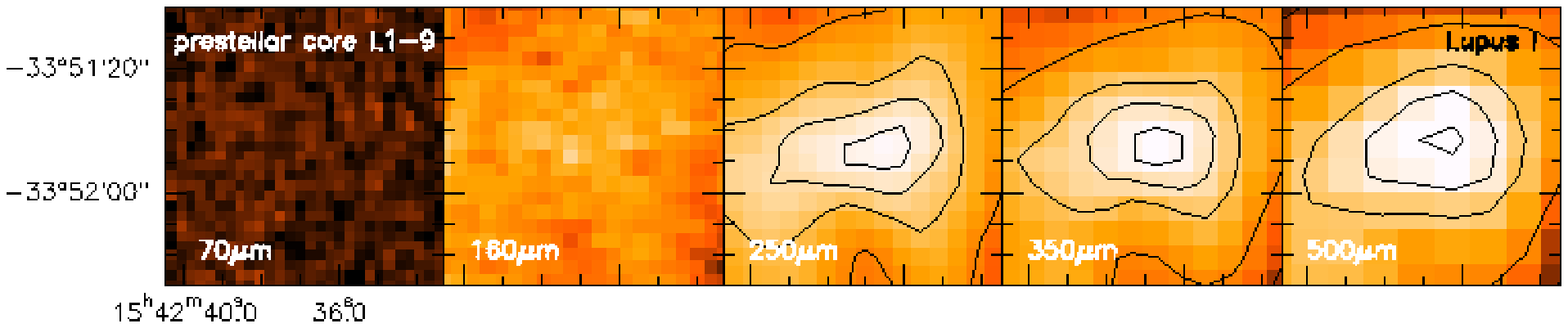}
\includegraphics[angle=-90,width=9cm]{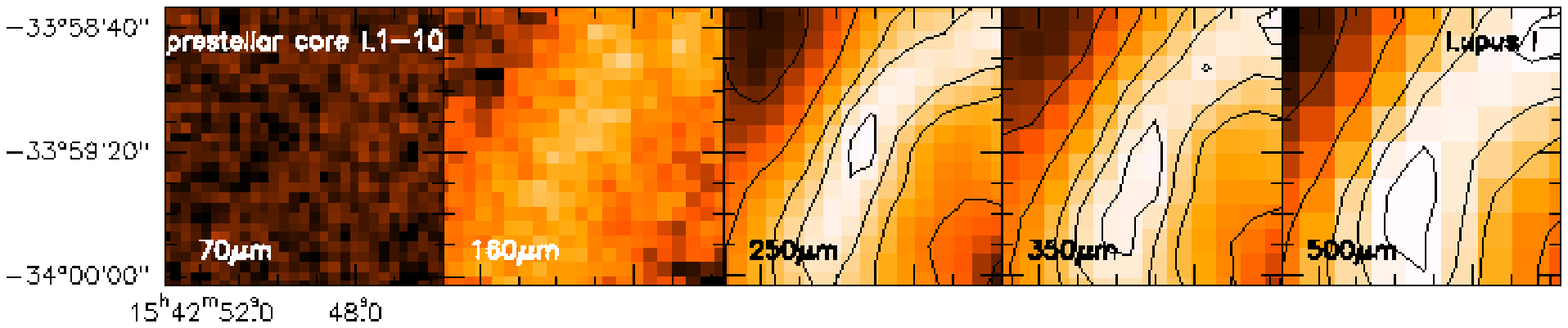}
\includegraphics[angle=-90,width=9cm]{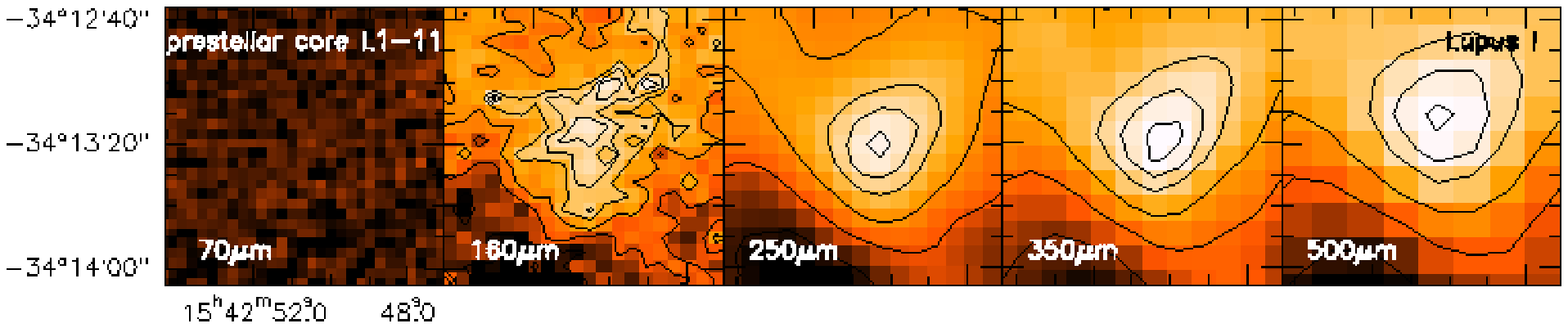}
\includegraphics[angle=-90,width=9cm]{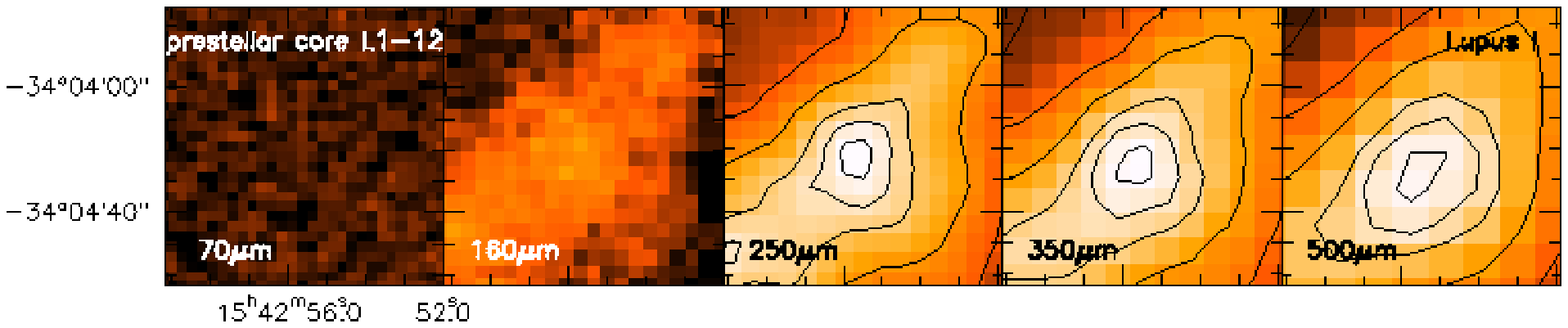}
\includegraphics[angle=-90,width=9cm]{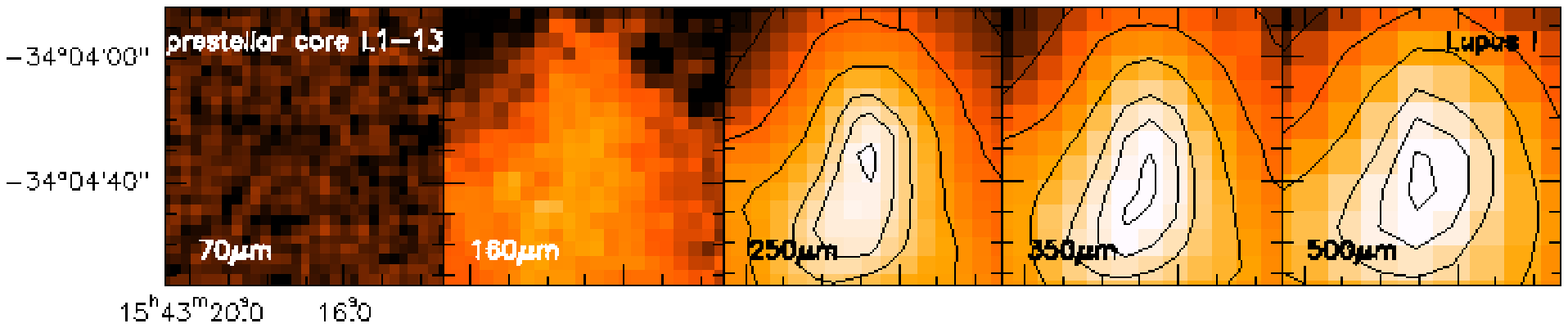}
\includegraphics[angle=-90,width=9cm]{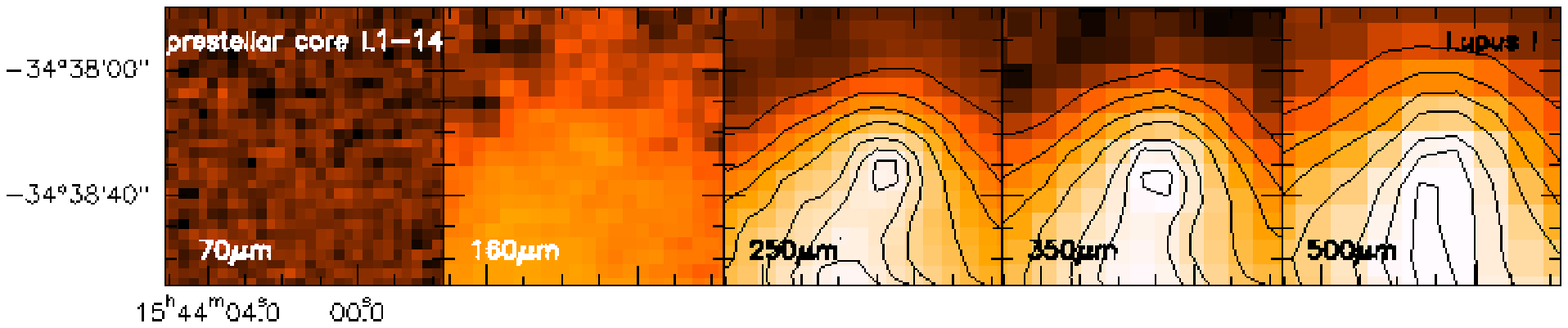}
\includegraphics[angle=-90,width=9cm]{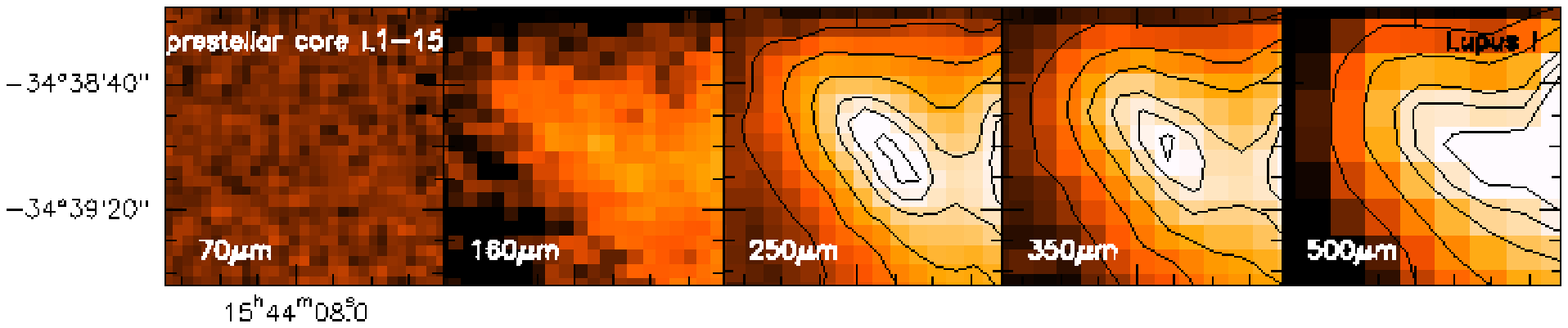}
\includegraphics[angle=-90,width=9cm]{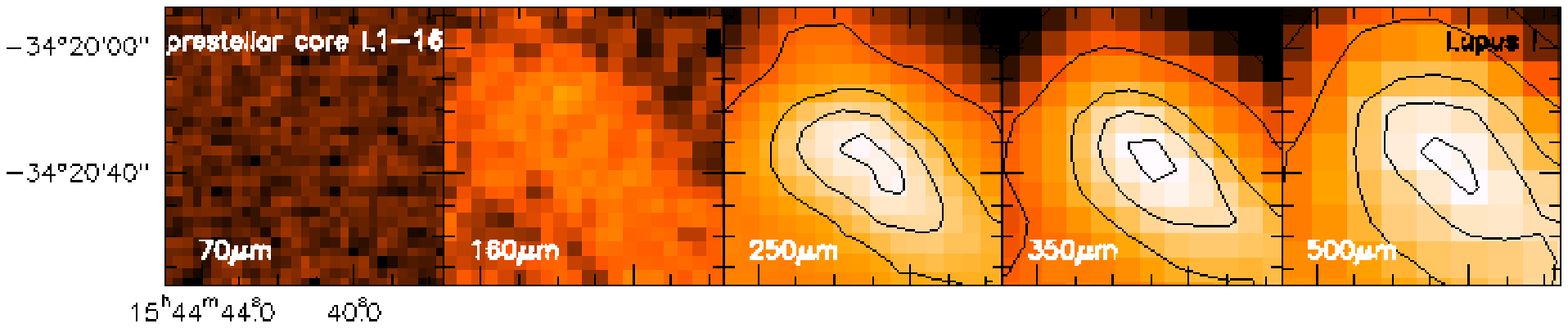}
\includegraphics[angle=-90,width=9cm]{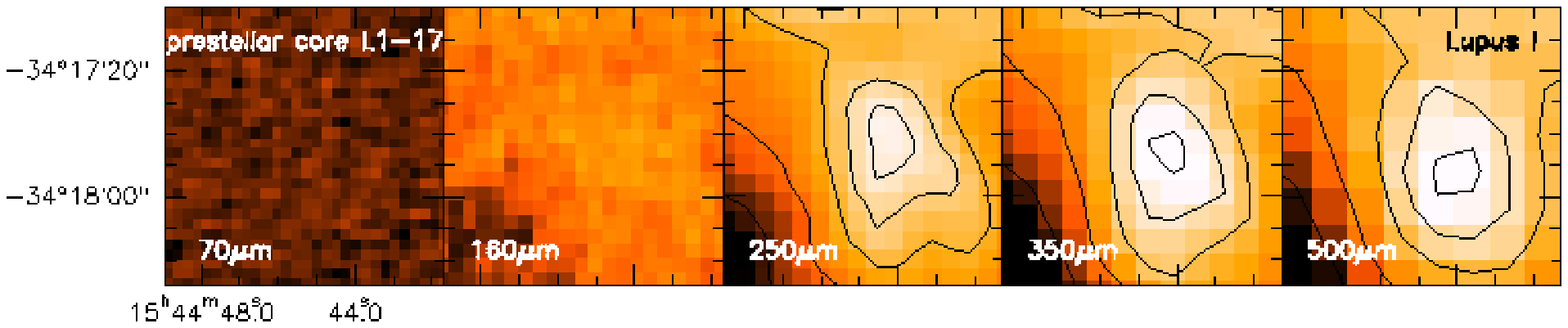}
\includegraphics[angle=-90,width=9cm]{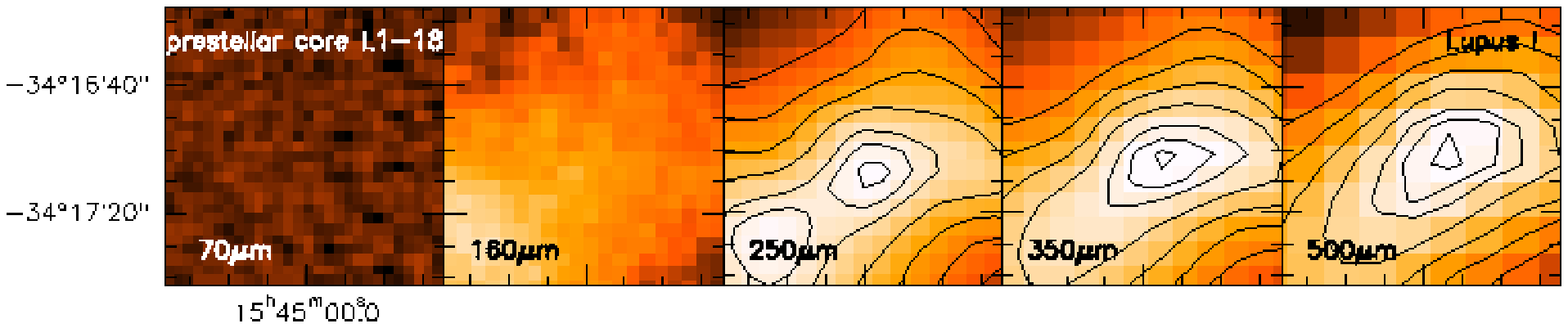}
\includegraphics[angle=-90,width=9cm]{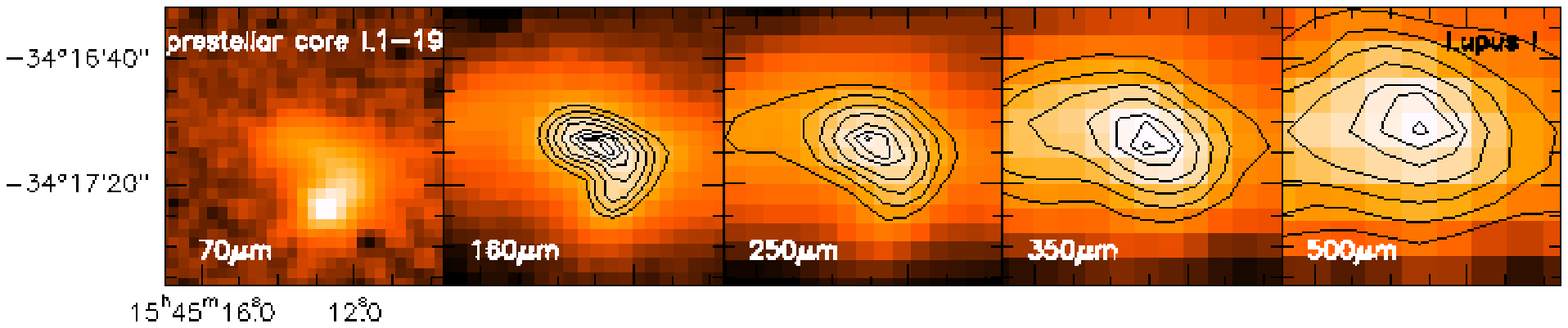}
\includegraphics[angle=-90,width=9cm]{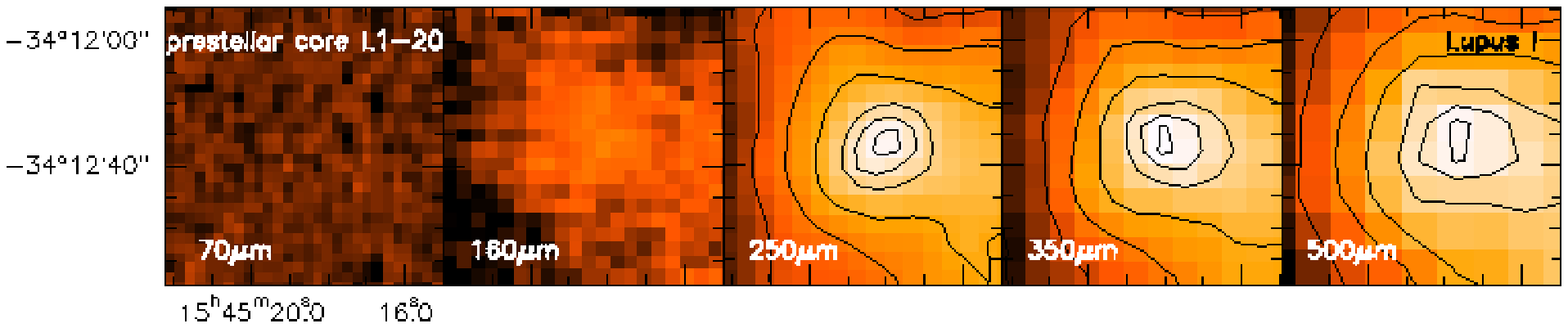}
\includegraphics[angle=-90,width=9cm]{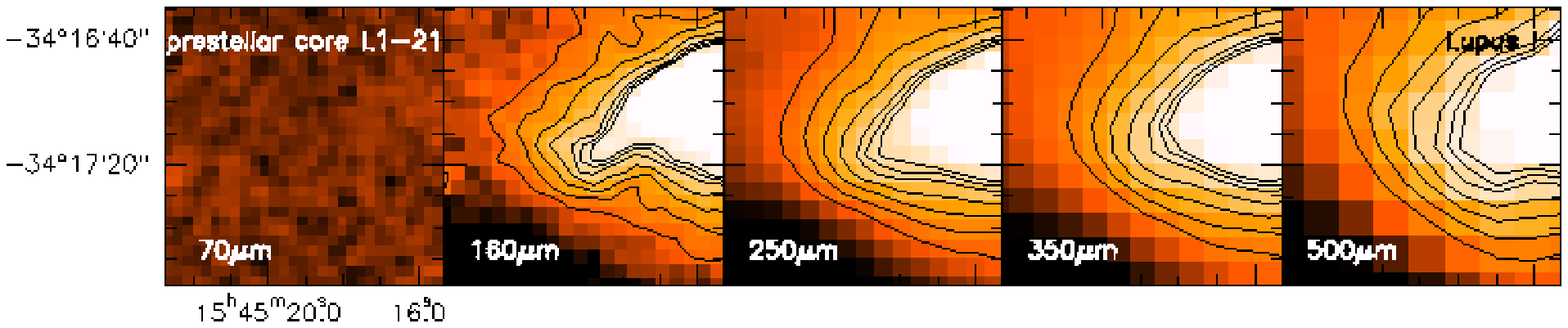}
\includegraphics[angle=-90,width=9cm]{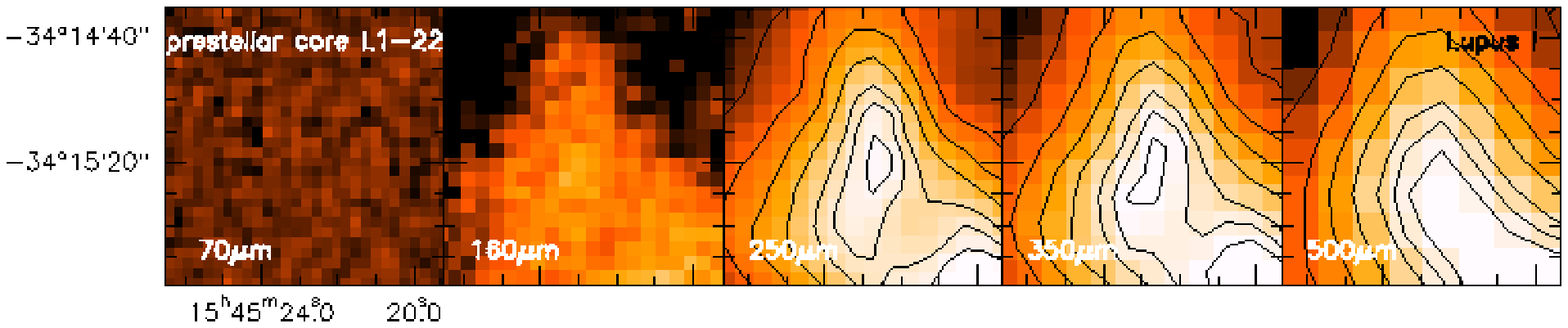}
\includegraphics[angle=-90,width=9cm]{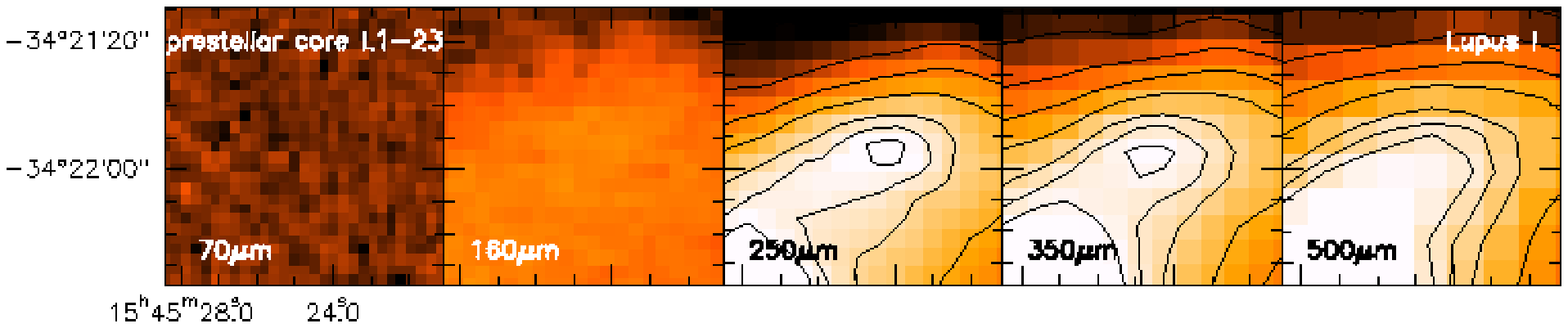}
\includegraphics[angle=-90,width=9cm]{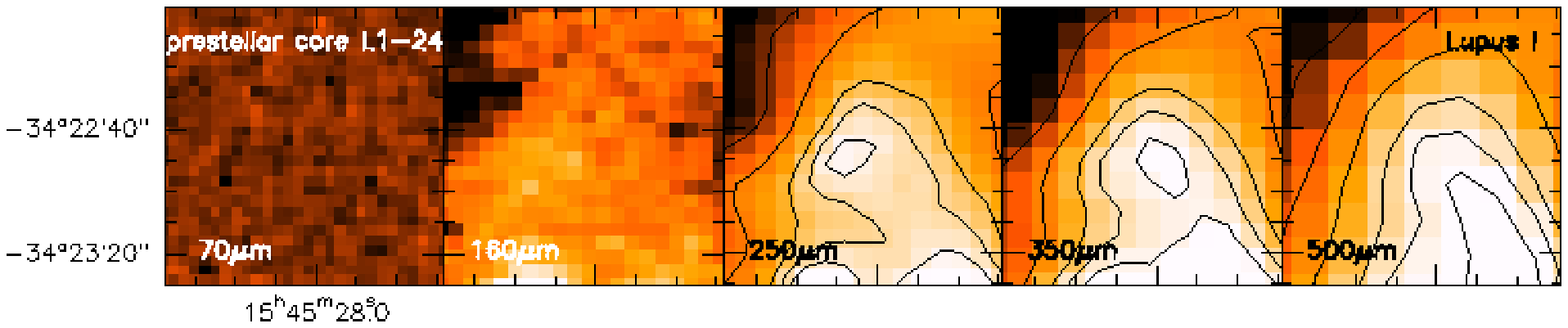}
\includegraphics[angle=-90,width=9cm]{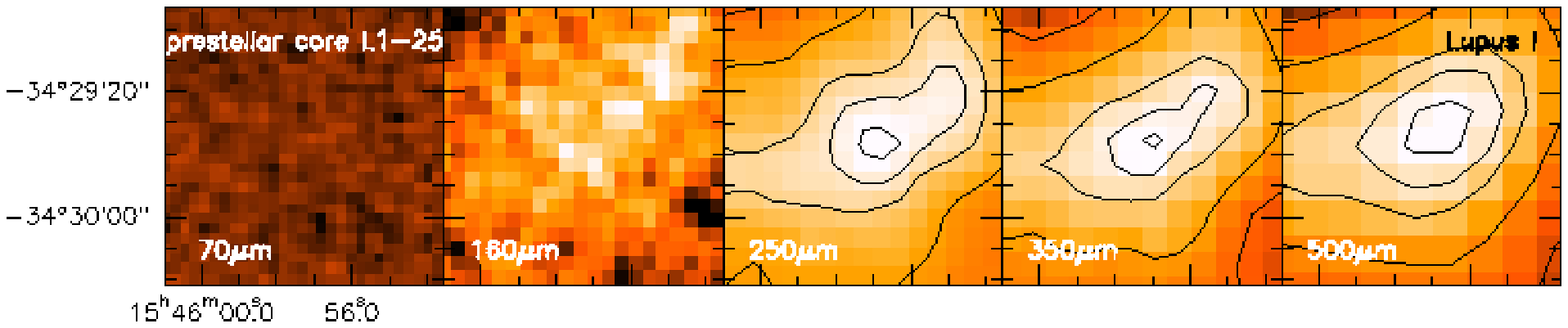}
\includegraphics[angle=-90,width=9cm]{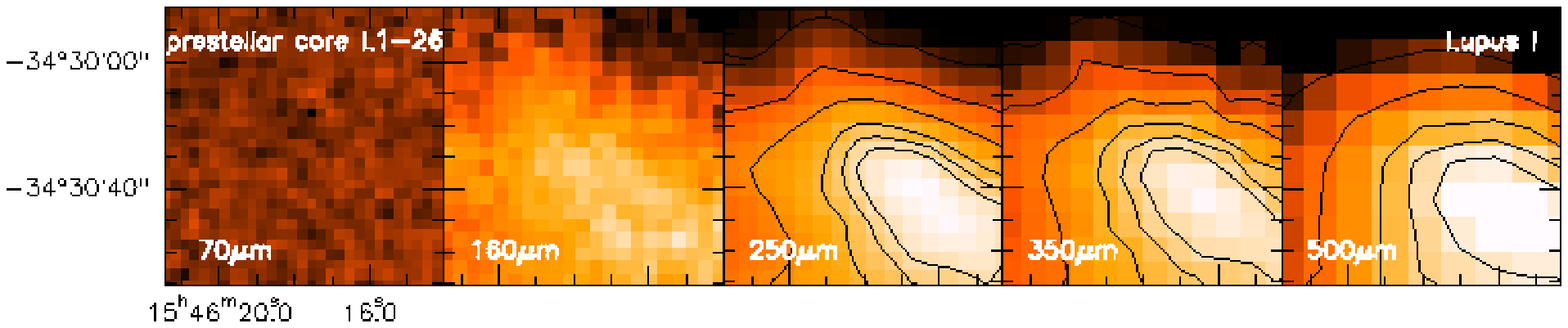}
\includegraphics[angle=-90,width=9cm]{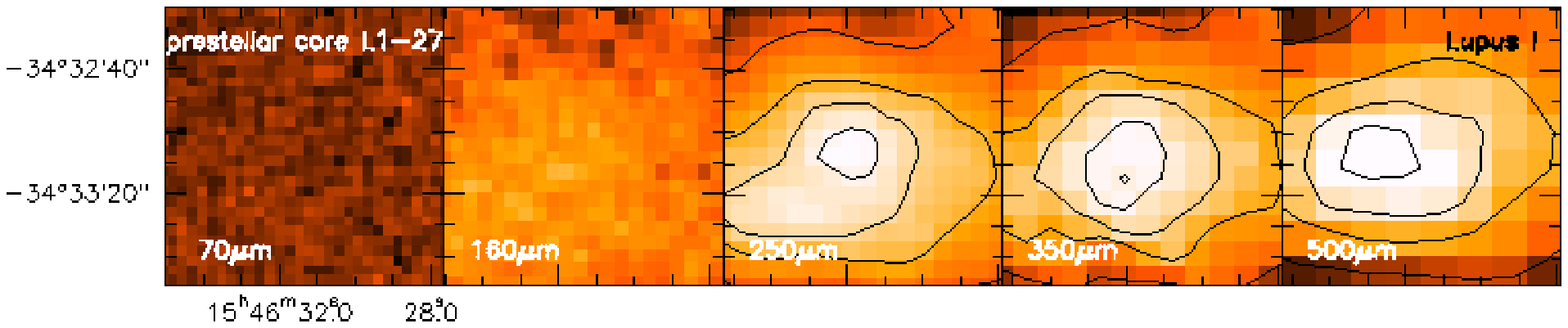}
\includegraphics[angle=-90,width=9cm]{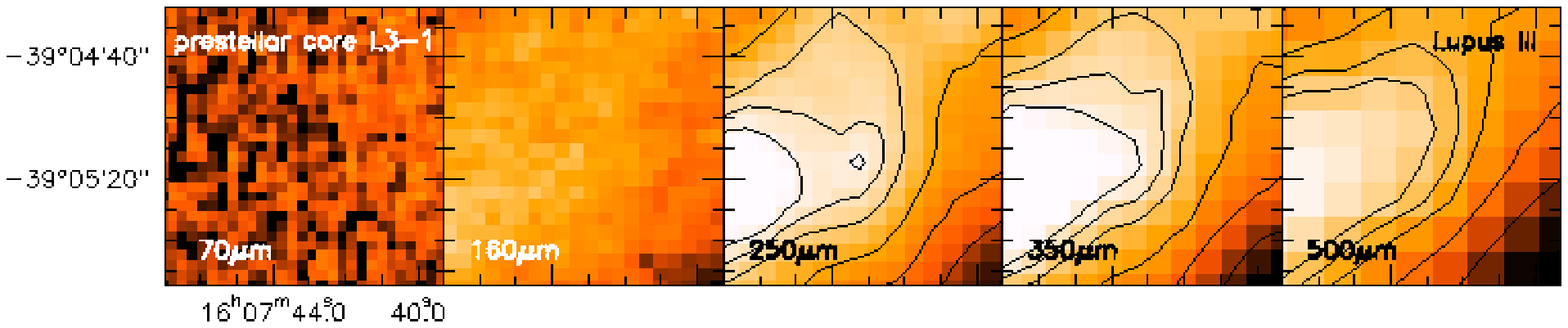}
\includegraphics[angle=-90,width=9cm]{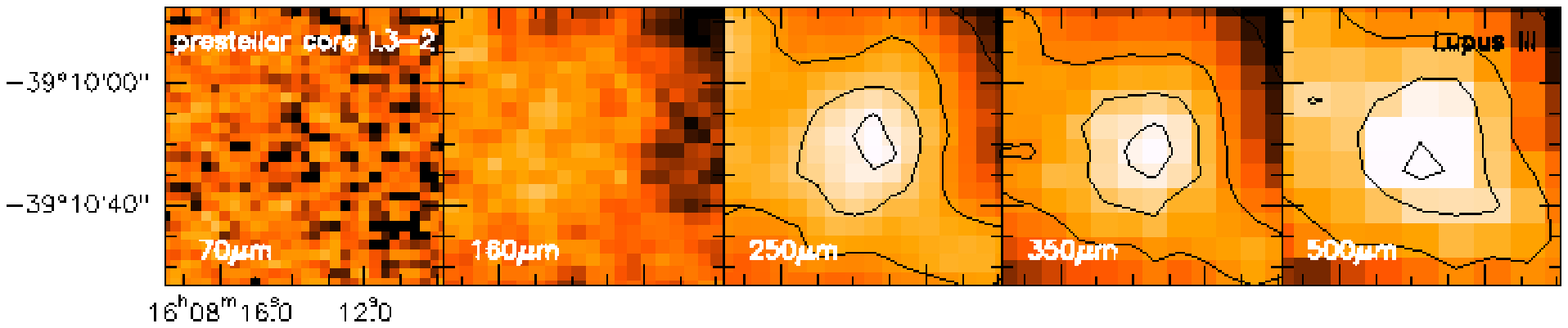}
\caption{\label{fig:youngsources} Visual catalog of the prestellar cores (contour levels are peakflux$\ast$0.99, 0.95, 0.90, 0.80, 0.70, 0.60, and 0.50) and Class 0 objects in the three Lupus clouds (I, III, IV), continued from Fig.\ref{fig:map5}. The maps are centered on the prestellar core/Class 0 object and are ordered per cloud by right ascension.}
\end{figure*}
\addtocounter{figure}{-1}
\begin{figure*}\centering
\includegraphics[angle=-90,width=9cm]{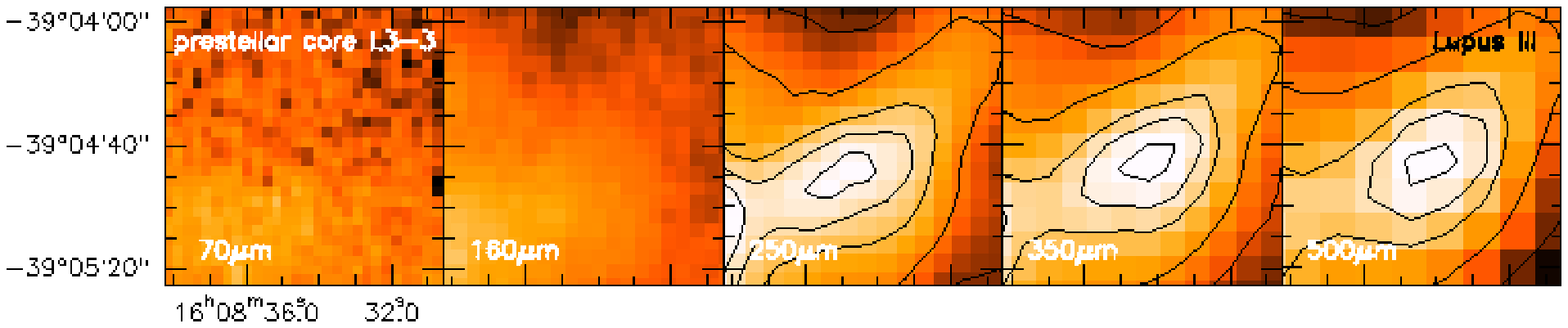}
\includegraphics[angle=-90,width=9cm]{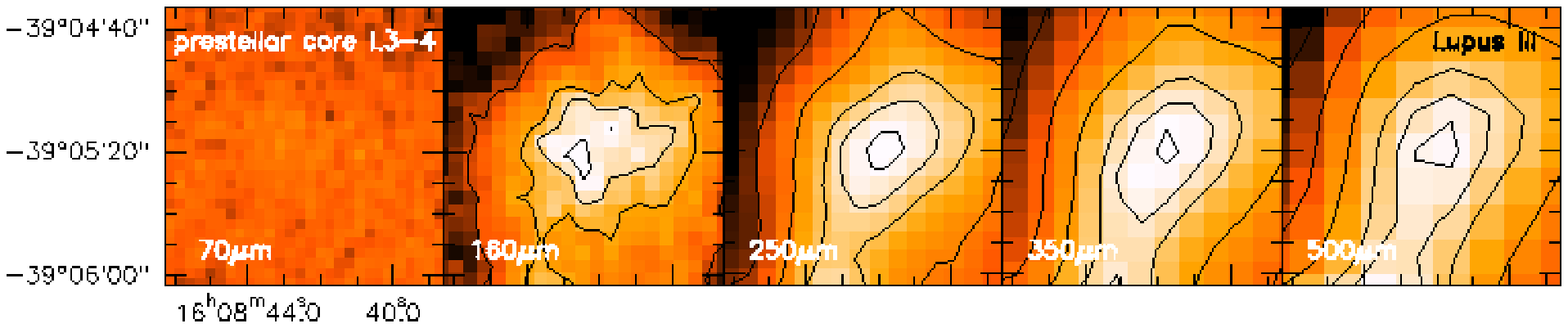}
\includegraphics[angle=-90,width=9cm]{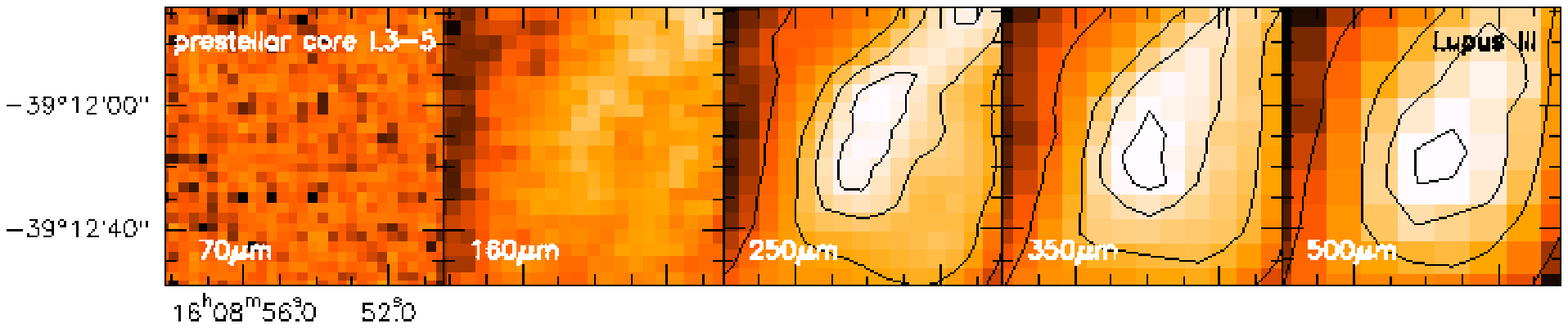}
\includegraphics[angle=-90,width=9cm]{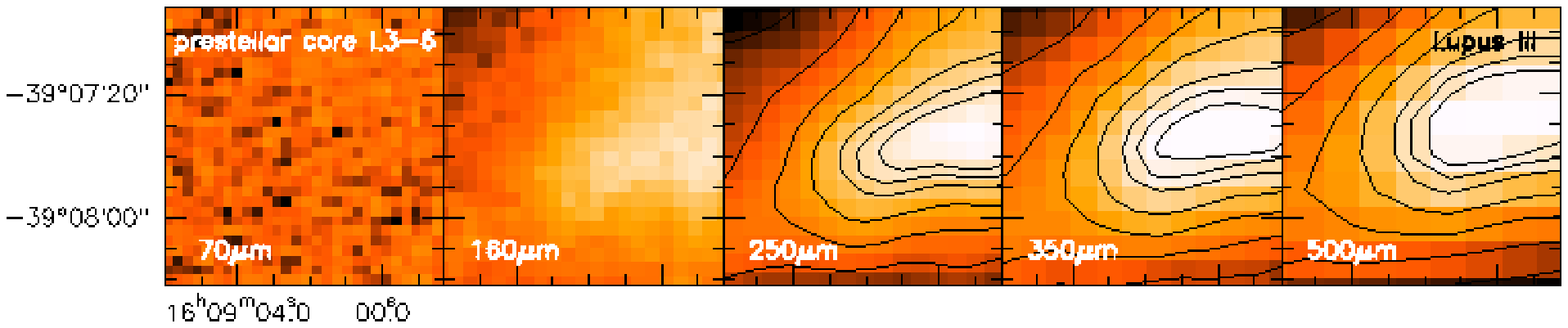}
\includegraphics[angle=-90,width=9cm]{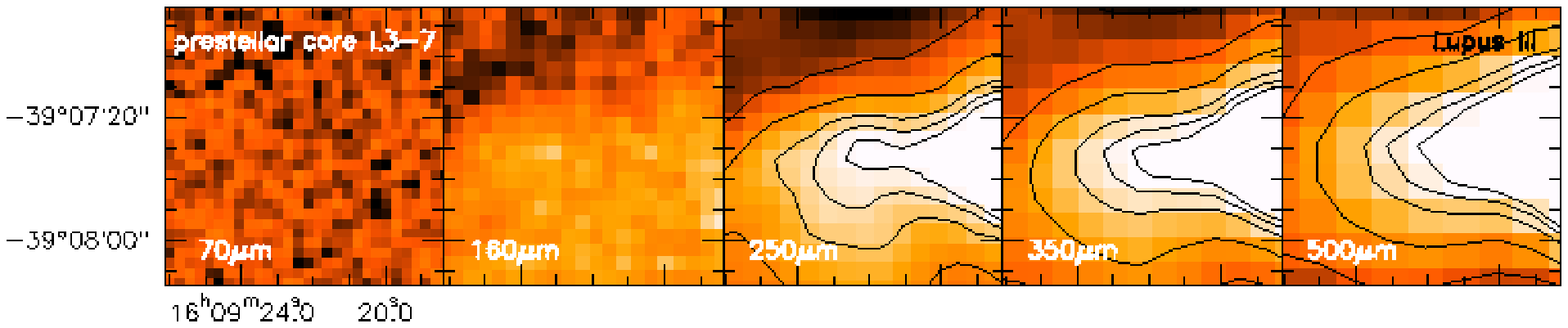}
\includegraphics[angle=-90,width=9cm]{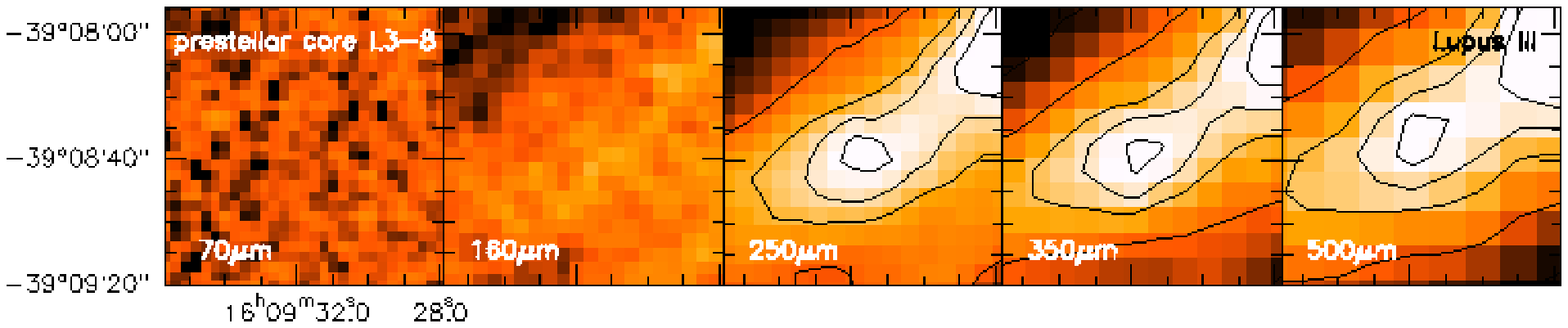}
\includegraphics[angle=-90,width=9cm]{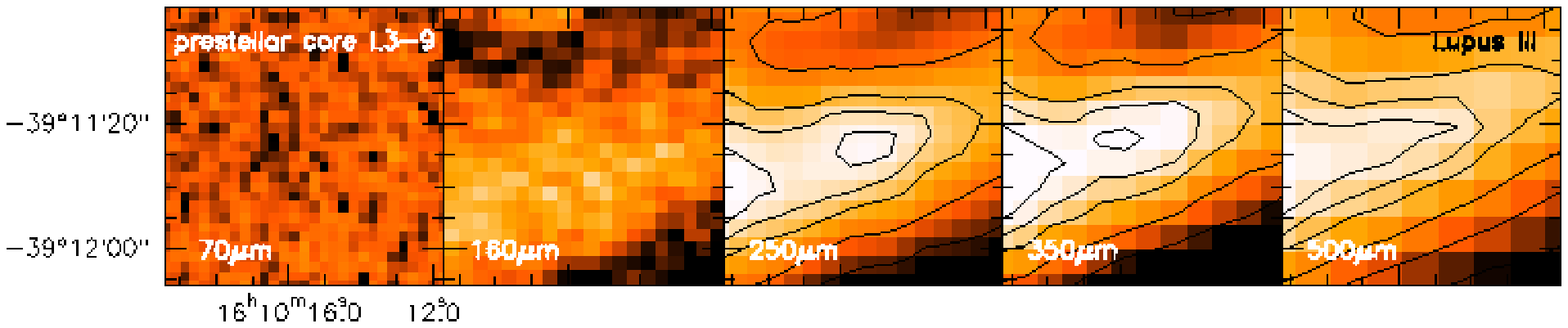}
\includegraphics[angle=-90,width=9cm]{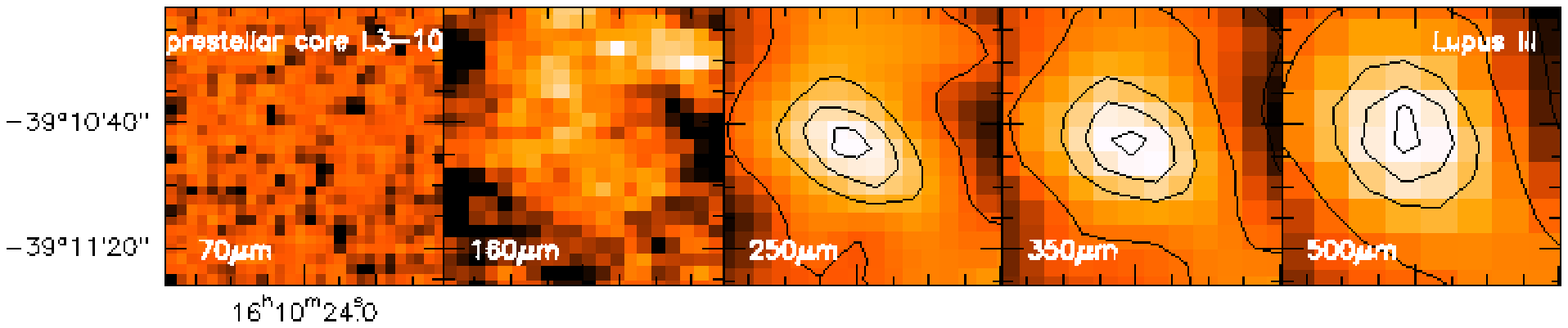}
\includegraphics[angle=-90,width=9cm]{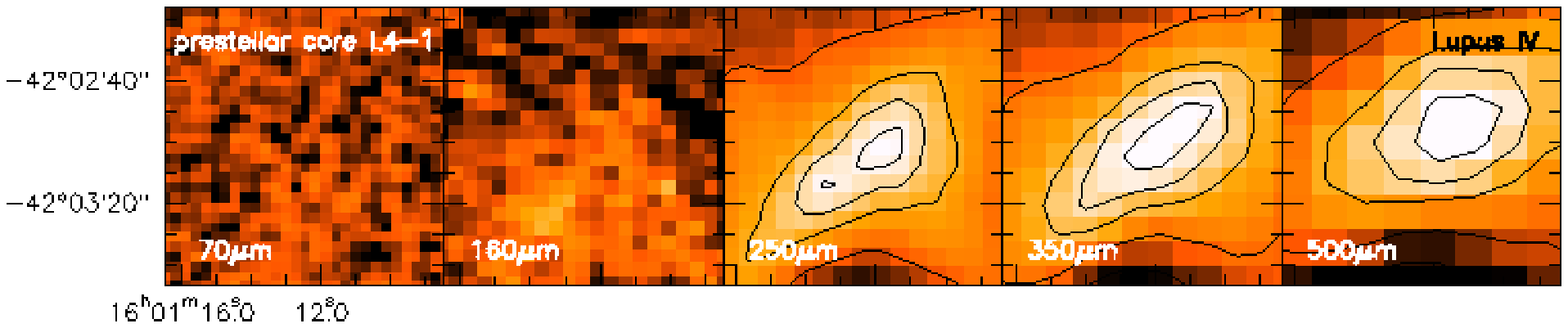}
\includegraphics[angle=-90,width=9cm]{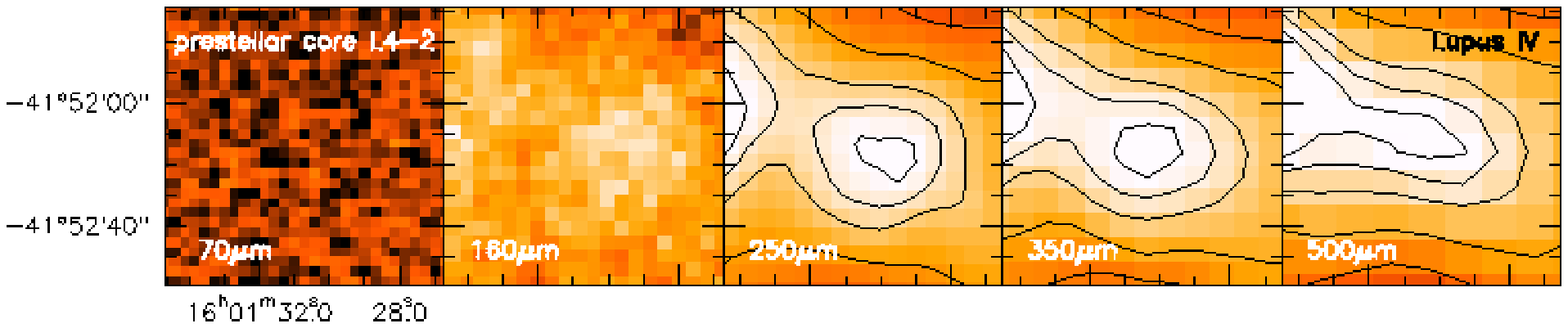}
\includegraphics[angle=-90,width=9cm]{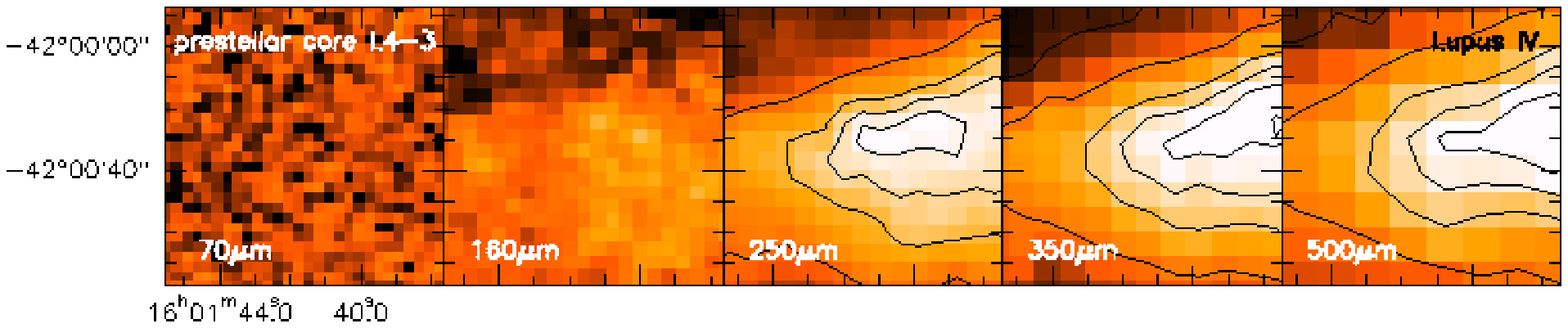}
\includegraphics[angle=-90,width=9cm]{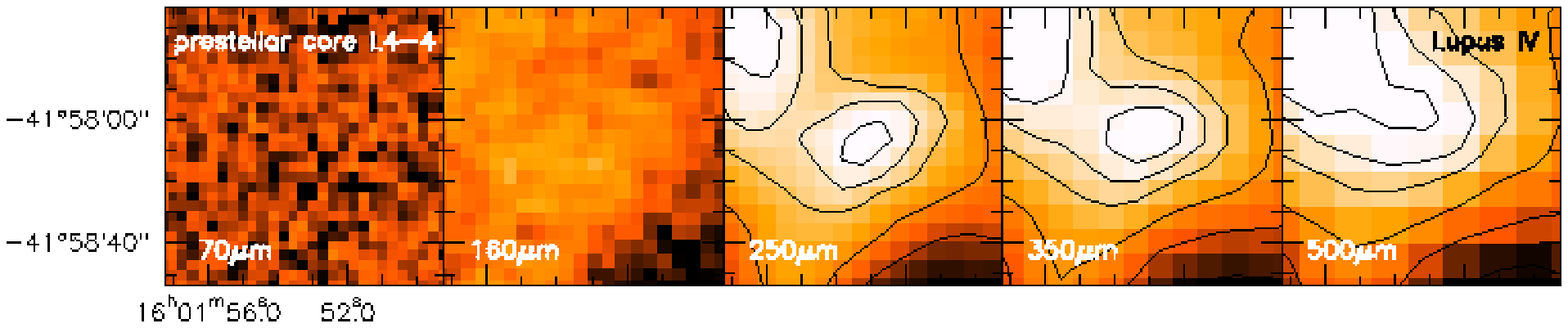}
\includegraphics[angle=-90,width=9cm]{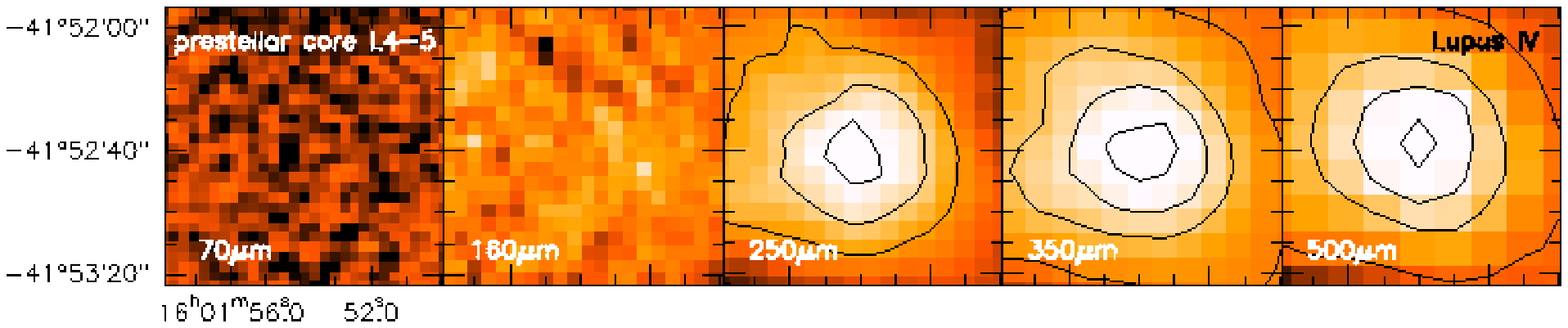}
\includegraphics[angle=-90,width=9cm]{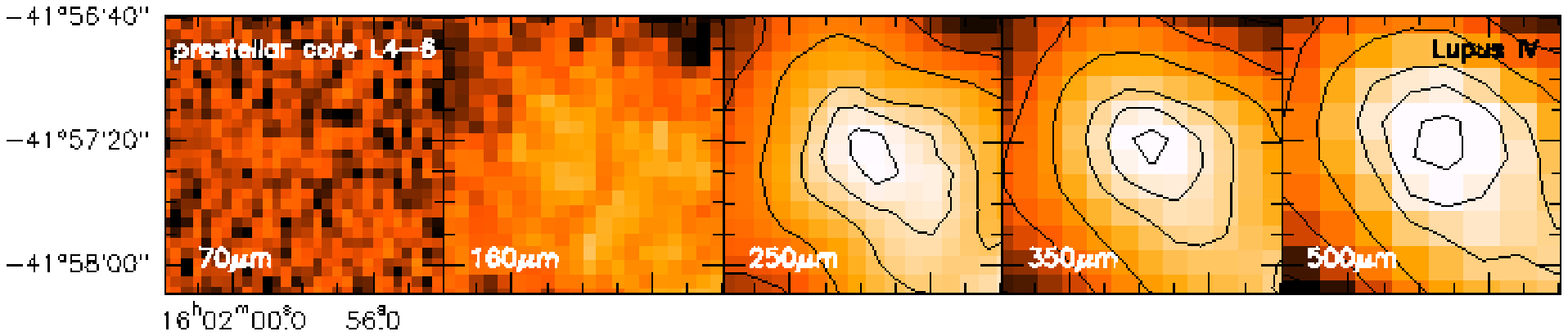}
\includegraphics[angle=-90,width=9cm]{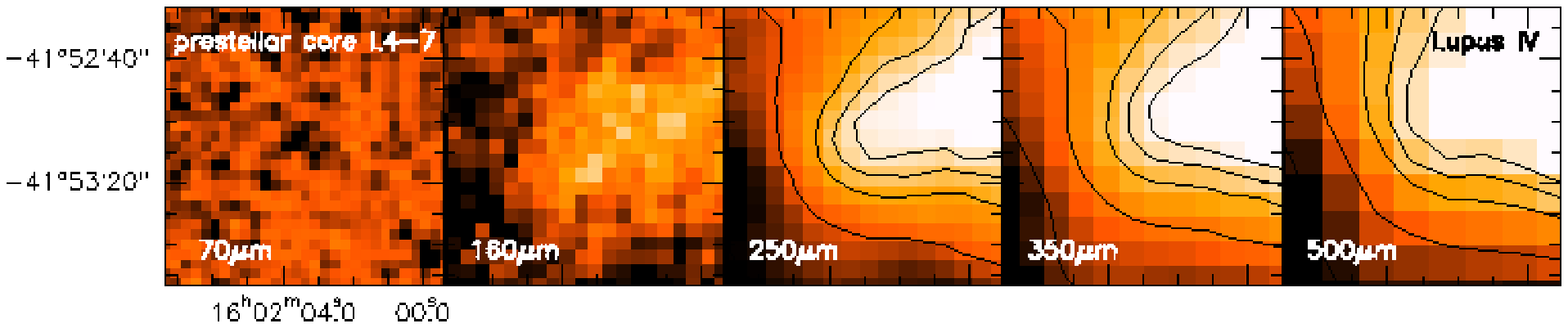}
\includegraphics[angle=-90,width=9cm]{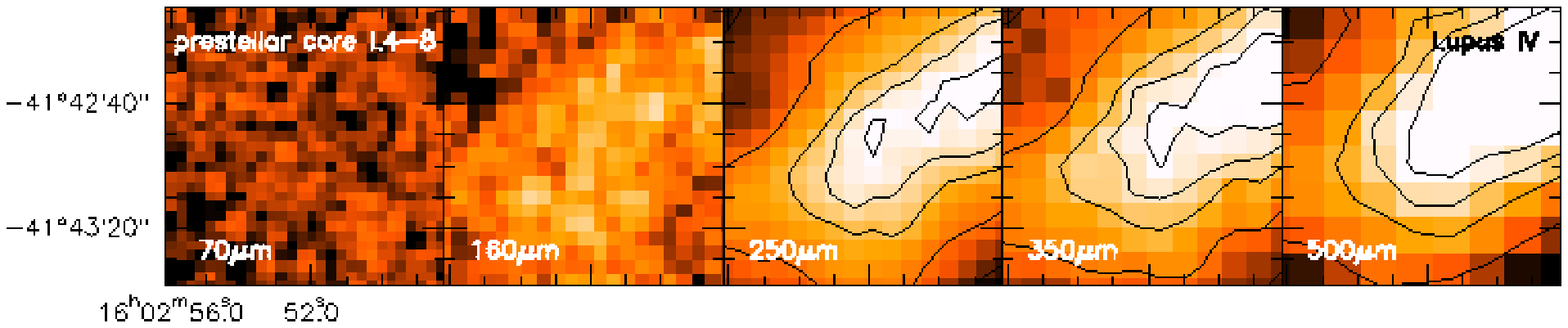}
\includegraphics[angle=-90,width=9cm]{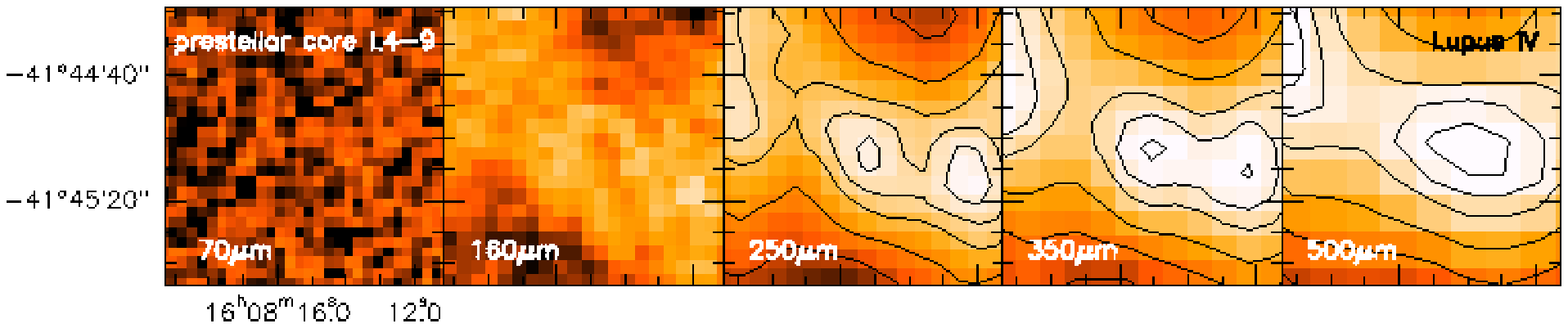}
\includegraphics[angle=-90,width=9cm]{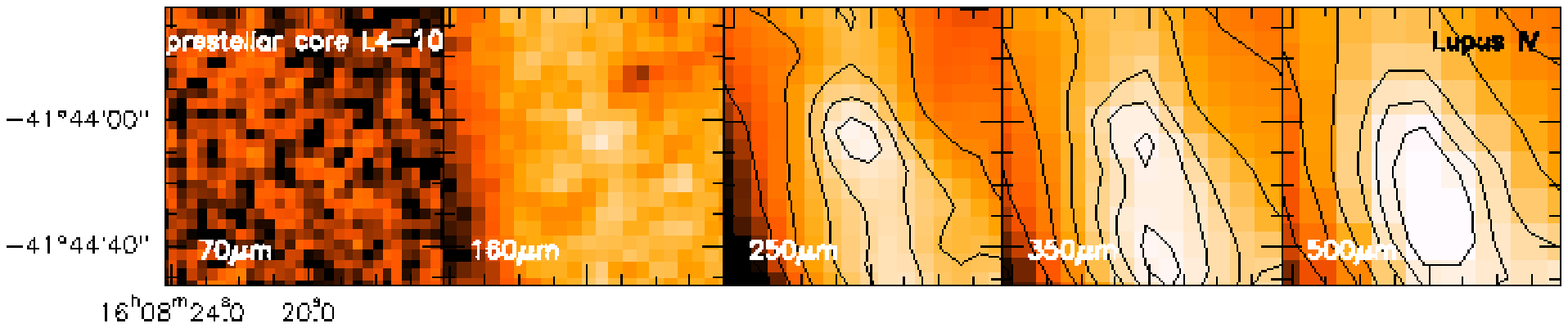}
\includegraphics[angle=-90,width=9cm]{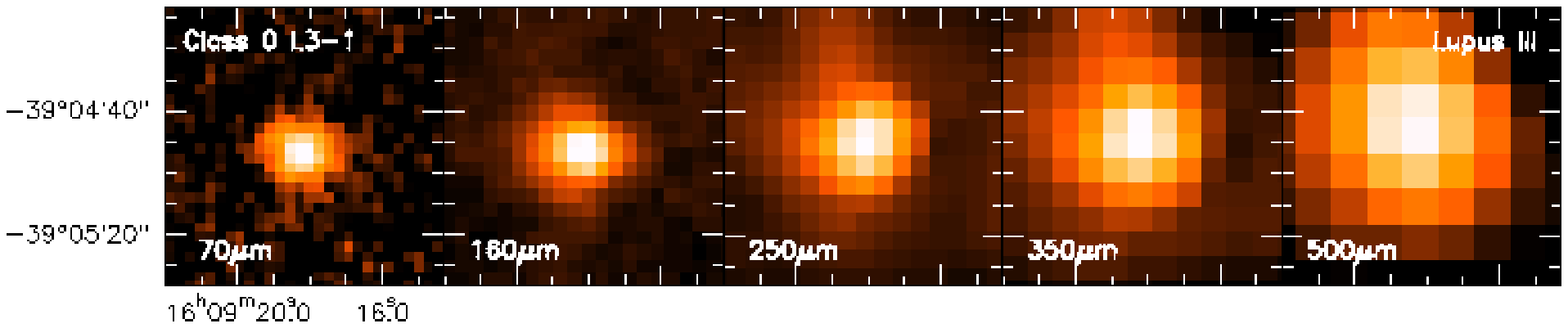}
\includegraphics[angle=-90,width=9cm]{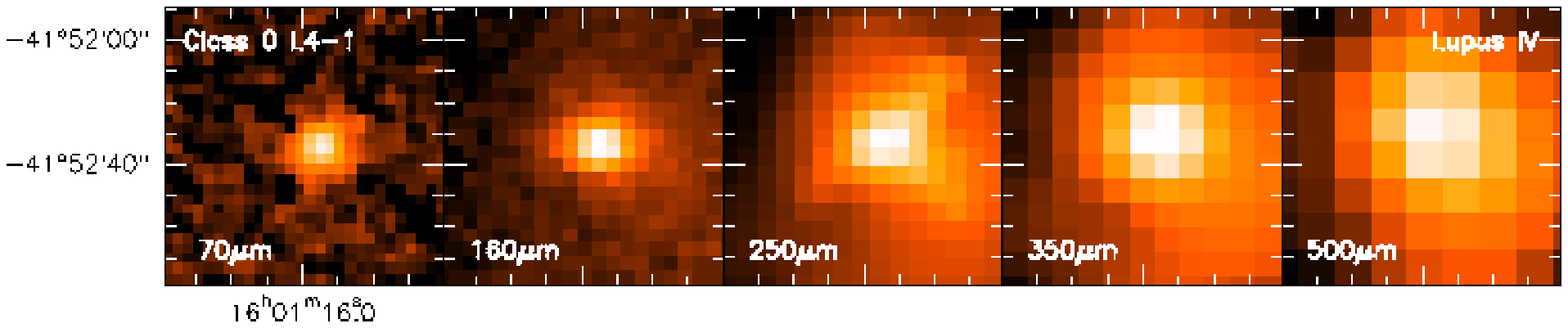}
\includegraphics[angle=-90,width=9cm]{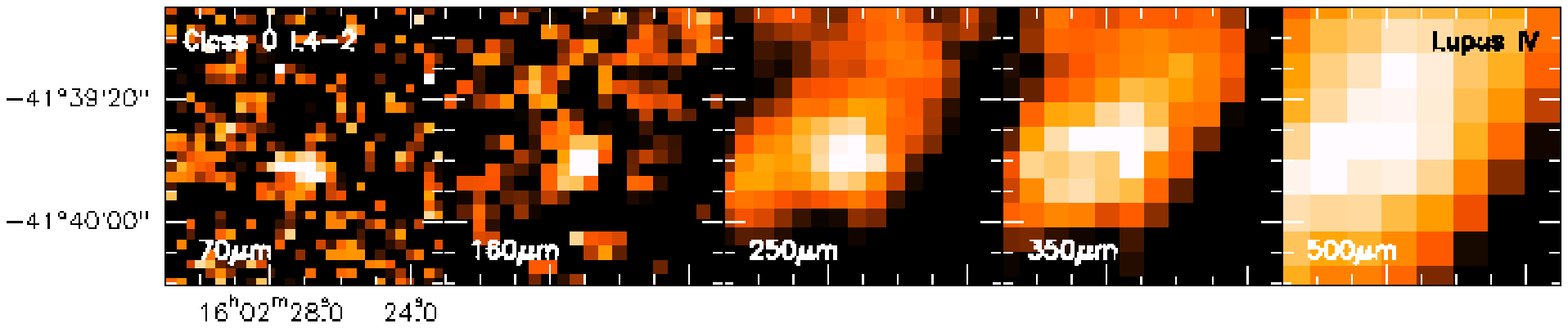}
\includegraphics[angle=-90,width=9cm]{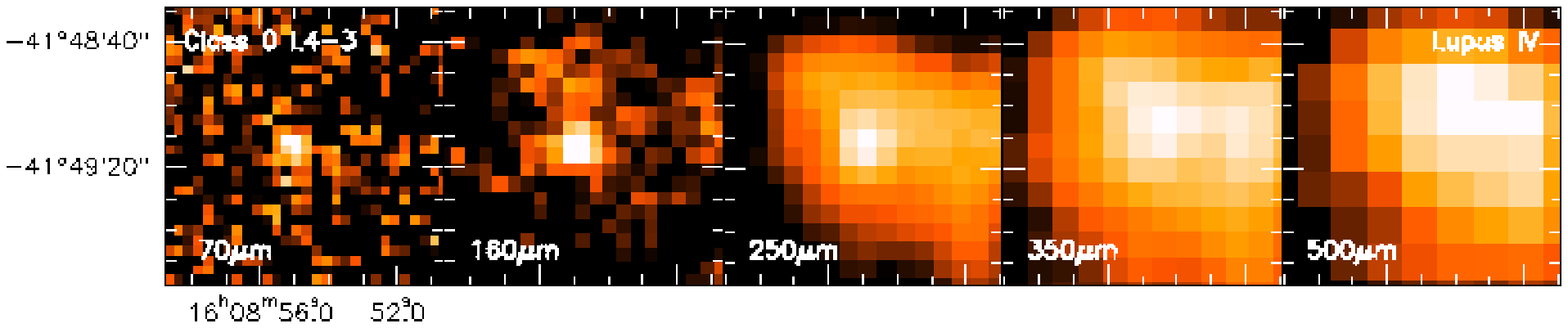}
\caption{-- {\em Continued}}
\end{figure*}
}

\end{document}